\documentclass[traditabstract]{aa}
\usepackage{ifpdf}
\usepackage{natbib}
\usepackage{times}
\usepackage{amssymb}
\usepackage{txfonts}
\ifpdf
  \usepackage[pdftex]{graphicx,color}
\else
  \usepackage[dvips]{graphicx,color}
\fi
\bibpunct{(}{)}{;}{a}{}{,}

\def\5{{\sc{v}}}
\def\4{{\sc{iv}}}
\def\3{{\sc{iii}}}
\def\2{{\sc{ii}}}
\def\1{{\sc{i}}}
\def\lam{{$\lambda$}}

\def\pA{$\phantom{1}$}
\def\kms{{\ensuremath{\mbox{km}\;\mbox{s}^{-1}}}}
\def\pmin{$\phantom{-}$}
\def\Teff{$T_{\rm eff}$}
\def\logg{$\log \, g$}
\def\logz{$\log[Z]$}
\def\chisq{$\chi^{2}$}

\begin{document}

\title{Stellar metallicities beyond the Local Group: the potential of $J$-band spectroscopy with extremely large telescopes}

\author{C.~J.~Evans\inst{1}, B.~Davies\inst{2,3,4}, R.-P.~Kudritzki\inst{5,6}
M.~Puech\inst{7}, Y.~Yang\inst{7,8}, J.-G.~Cuby\inst{9}\\
D.~F.~Figer\inst{2}, M.~D.~Lehnert\inst{7}, S.~L.~Morris\inst{10} \and G. Rousset\inst{11}}

\offprints{C.~J.~Evans at chris.evans@stfc.ac.uk}

\authorrunning{C.~J.~Evans et al.}

\titlerunning{Stellar metallicities from $J$-band spectroscopy with ELTs}

\institute{UK Astronomy Technology Centre, Royal Observatory Edinburgh, Blackford Hill, Edinburgh, EH9 3HJ, UK
\and 
Rochester Institute of Technology, 54 Lomb Memorial Drive, Rochester, NY 14623, USA
\and 
School of Physics \& Astronomy, University of Leeds, Woodhouse Lane, Leeds, LS2 9JT, UK
\and 
Institute of Astronomy, University of Cambridge, Madingley Road, Cambridge, CB3 0HA, UK
\and 
Institute for Astronomy, University of Hawaii, 2680 Woodlawn Drive, Honolulu, HI 96822, USA
\and 
Max-Planck-Institute for Astrophysics, Karl-Schwarzschild-Str. 1, 85748 Garching, Germany
\and 
GEPI, Observatoire de Paris, 5 Place Jules Janssen, 92195 Meudon Cedex, France
\and 
Key Laboratory of Optical Astronomy, National Astronomical Observatories, Chinese Academy of 
Sciences, 20A Datun Road, Chaoyang District, Beijing 100012, China
\and 
LAM, OAMP, 38 rue Fr\'{e}d\'{e}ric Joliot Curie, 13388 Marseille Cedex 13, France
\and 
Department of Physics, Durham University, South Road, Durham, DH1~3LE, UK
\and 
LESIA, Observatoire de Paris, 5 Place Jules Janssen, 92195 Meudon Cedex, France
}

\date{Accepted 8 December 2010}

\abstract{We present simulated $J$-band spectroscopy of red giants and
  supergiants with a 42\,m European Extremely Large Telescope (E-ELT),
  using tools developed toward the EAGLE Phase A instrument study.
  The simulated spectra are used to demonstrate the validity of the
  1.15-1.22\,$\mu$m region to recover accurate stellar metallicities
  from Solar and metal-poor (one tenth Solar) spectral templates.
  From tests at spectral resolving powers of four and ten thousand, we
  require continuum signal-to-noise ratios in excess of 50 (per
  two-pixel resolution element) to recover the input metallicity to
  within 0.1\,dex.  We highlight the potential of direct estimates of
  stellar metallicites (over the range $-$1\,$<$\,[Fe/H]\,$<$\,0) of
  red giants with the E-ELT, reaching out to distances of $\sim$5\,Mpc
  for stars near the tip of the red giant branch.  The same
  simulations are also used to illustrate the potential for
  quantitative spectroscopy of red supergiants beyond the Local Volume
  to tens of Mpc.  Calcium triplet observations in the $I$-band are
  also simulated to provide a comparison with contemporary techniques.
  Assuming the EAGLE instrument parameters and simulated performances
  from adaptive optics, the $J$-band method is more sensitive in terms
  of recovering metallicity estimates for a given target.  This
  appears very promising for ELT studies of red giants and
  supergiants, offering a {\em direct} metallicity tracer at a
  wavelength which is less afffected by extinction than shortward
  diagnostics and, via adaptive optics, with better image quality.}


\maketitle

\section{Introduction}\label{intro} 

Plans are well advanced for the next generation of optical and
infrared (IR) ground-based telescopes, the extremely large telescopes
(ELTs).  Their science cases are broad, ranging from direct imaging of exoplanets, to
studies of resolved stellar populations in external galaxies, and
spectroscopy of distant `first light' galaxies at the highest
redshifts \citep[e.g.][]{hdg06}.  

The ELTs are an increasingly global effort, with three projects
under detailed study --  the Giant Magellan Telescope (GMT), the
Thirty Meter Telescope (TMT), and the European Extremely Large
Telescope (E-ELT).  With primary apertures in excess of 20\,m, they will
deliver a huge gain in our capabilities via a combination
of unprecedented sensitivity and exquisite angular resolution.
In parallel to the design of the observatories,
significant effort has also been invested in studies of ELT instrumentation
\citep[see][]{j10,s10,skr10}.

Over the past few years, deep imaging from ground-based telescopes and
the {\em Hubble Space Telescope (HST)} has provided us with new and
unique views of the outer regions of large galaxies beyond the Milky
Way for the first time, such as M31 \citep[e.g.][]{f05} and M33
\citep[e.g.][]{bark07}. From analysis of the resulting photometry and
colour-magnitude diagrams we can determine star-formation and assembly
histories for external galaxies, enabling tests of theoretical models
of galaxy evolution \citep[e.g.][]{bj05}, i.e. using resolved stellar
populations as a tracer of the processes which have shaped the
evolution of their host systems.

The imaging from the Advanced Camera for Surveys (ACS) Nearby Galaxy
Survey Treasury \citep[ANGST;][]{angst} provides an excellent
illustration of the diversity of galaxies beyond the Local Group.
Crucially, the Local Volume includes a wide variety of morphological
types -- massive ellipticals, large metal-poor irregulars, lower-mass
late-type spirals, interacting systems, dwarf starbursts -- providing
an excellent opportunity to quantify the effects of environment on
galaxy evolution.

Photometric methods are powerful when applied to extragalactic stellar
populations but follow-up spectroscopy can aid our understanding
significantly via precise chemical abundances and stellar kinematics.
For example, results from the spectroscopy of luminous blue
supergiants in external galaxies obtained by the Araucaria project
\citep{araucaria}.  However, in pursuit of spectroscopy of evolved
stellar populations, the 8-10m class telescopes are already near their
limits beyond a few hundred kpc. For instance, observations with
Keck-DEIMOS struggled to yield useful spectra below the tip of
the red giant branch (TRGB) in M31 at $I$~$>$~21.5 \citep{c06}.  If we aspire
to spectroscopy of individual evolved stars in galaxies beyond the
Local Group, we require the increased sensitivity of the ELTs, likely
combined with some degree of correction from adaptive optics (AO) to
mitigate the effects of crowding.

Indeed, for stellar spectroscopy with the ELTs there will be a fine
balance in sensitivity between the improved image quality from AO as
one goes to longer wavelengths (where the wavefront errors become less
significant compared to the observed wavelengths) versus the increased
sky background.  Moreover, work over the past decades has provided an
excellent understanding of many of the optical spectral lines
available to us, enabling robust estimates of physical parameters and
chemical abundances. To exploit the best performance (in terms of
angular resolution) from the ELTs, we need to improve our knowledge of
near-IR diagnostics, and also to pin-down the `sweet spot' in terms of
the gain in sensitivity from AO versus the background contribution.

A recent study by Davies, Kudritzki \& Figer (\citeyear{dkf10};
hereafter DKF10) suggested new $J$-band diagnostics, spanning
1.15-1.22\,$\mu$m (including lines from Mg, Si, Ti, and Fe), as a
means to obtain stellar metallicities in extra-galactic red
supergiants (RSGs). This wavelength region is relatively unexplored in
this regard, only previously considered by \citet{o04}.  DKF10 noted
that this method could be very powerful with new IR instruments under
construction for 8-10\,m class telescopes (such as VLT-KMOS and
Keck-MOSFIRE), providing precise stellar abundances in galaxies out to
a potential distance of $\sim$10\,Mpc, complementing those from
luminous blue supergiants \citep[e.g.][]{b01,kud08}.  An even more
compelling future prospect is in the context of ELT observations, with
DKF10 noting the potential of direct abundance estimates for
individual RSGs in galaxies even beyond the Local Volume (subject to
crowding).

A second possible application of this spectral region is in
observations of evolved red giant branch (RGB) stars -- the long-lived
descendants of much lower-mass stars.  The {\sc marcs} model
atmospheres \citep{g08} used by DKF10 are actually of red giants (see
their Section~2.2 for a discussion of the applicability of these
models over a large range of stellar luminosities). 

The potential of a direct abundance diagnostic for $J$-band ELT
observations of both young (RSG) and old (RGB) stellar populations in
external galaxies -- with the benefit of better AO correction compared
to shorter wavelengths -- warrants further exploration.  EAGLE is a
conceptual design of an AO-corrected, multi-IFU, near-IR spectrograph,
undertaken as one of the Phase~A instrument studies for the E-ELT
\citep{cuby10}.  In this article we employ tools developed in the
course of the EAGLE study as a proof-of-concept for quantitative
$J$-band spectroscopy with the ELTs.

We simulate EAGLE $J$-band observations with two objectives.  First,
to validate the technique of DKF10 for metal-poor spectral templates,
and secondly, to explore the distances to which we could obtain robust
abundance estimates (for both RGB stars and RSGs) with ELT
observations.  To compare these results with diagnostics already in
common use, we also consider simulations in the $I$ band.
Section~\ref{models} describes the tools and assumptions used in the
simulations, with the analysis presented in Section~\ref{analysis}.
In Sections~\ref{discussion} and \ref{rsgs} we discuss the results in
the context of the scientific potential of the E-ELT and other
upcoming facilities.

\section{Simulations}\label{models}

Three top-level science cases informed the requirements
of the EAGLE design: the physics and evolution of high-redshift
galaxies, characterization of `first light' galaxies at the highest
redshifts, and studies of resolved stellar populations in the Local
Volume.  The specifications of the baseline design are summarised in
Table~\ref{specs} (with further discussion given by \citeauthor{e10} 
2010b).

To investigate the critical requirements for spatially-resolved
spectroscopy of high-redshift galaxies, \citet{mp08} developed a
web-based tool to generate simulated IFU datacubes.  This was used by
\citeauthor{mp08} (2008, 2010a) to investigate the required image
quality, in terms of the level of AO correction, to recover
spatially-resolved properties in high-redshift galaxies (up to
$z$\,$\sim$\,6).  Functional details of the {\sc websim} simulator are
given by \citeauthor{mpspie} (2010b).  In brief, there is a
web-interface to a core {\sc idl} code. The user uploads input
datacubes in the Flexible Image Transport System (FITS) format, which
must incorporate specific keywords in the headers to provide
information to the code, as well as sufficient spatial sampling
(typically ten times greater than the size of the final IFU pixel) and
the desired spectral sampling.  The input cubes are then convolved
with a model point-spread function (PSF), selected from a list of AO
simulations (Section~\ref{moaopsfs}) which are hard-wired into the
code.  By adopting the relevant instrument/telescope parameters,
\citeauthor{mp10} (2010a) provided an external check of the {\sc
  websim} routines by reproducing datacubes to match those from
SINFONI observations of high-redshift galaxies.

\begin{table}
\begin{center}
\caption{EAGLE baseline design. The patrol field is the instrument field-of-view
within which integral-field units (IFUs) can be configured to observe
individual targets/sub-fields.}\label{specs}
\begin{tabular}{ll}
\hline\noalign{\smallskip}
Parameter & Specification \\
\noalign{\smallskip}\hline\noalign{\smallskip}
Patrol Field & Eqv. 7$^\prime$ diameter \\
IFU field-of-view & 1{\mbox{\ensuremath{.\!\!^{\prime\prime}}}}65 $\times$ 1{\mbox{\ensuremath{.\!\!^{\prime\prime}}}}65 \\
Multiplex (\# of IFUs) & 20 \\
Spatial resolution & 30\% EE in 75 mas ($H$ band)\\
Spectral resolving power ($R$) & 4000 \& 10000\\
Wavelength range & 0.8-2.5$\mu$m\\
\noalign{\smallskip}\hline
\end{tabular}
\end{center}
\end{table}

The parameters used in the {\sc websim} simulations are summarised in
Table~\ref{simspecs}.  The current E-ELT design features a primary
with an equivalent diameter of 42\,m and with approximately 9\%
obscuration by the secondary and supporting structures.  The telescope
throughput ($\ge$80\%) was the expected temporal average of the E-ELT
design at the start of the EAGLE study, retained here as a worst-case
estimate (compared to the $\ge$90\% adopted by the Design
Reference Mission studies, e.g., \citeauthor{mp10} 2010a).

Inclusion of the sky background is discussed in detail by
\citeauthor{mp10} (2010a), in which a well-sampled model of the Mauna Kea
background is scaled by measurements of the atmospheric absorption at
Paranal; the same approach is used here.  The E-ELT site was announced
in early 2010 to be Cerro Armazones in northern Chile. Although at an
altitude of 3060\,m, slightly higher than Paranal (the site of the
Very Large Telescope, VLT), the adopted sky model is a good enough
approximation for our purposes.

The {\sc websim} routines add the sky spectrum (which includes the OH
emission lines) to each spatial element of the datacube, then a noise
contribution is added. In parallel, a separate sky spectrum is
constructed using a different realisation of the noise (i.e. a
different random number seed), to simulate an observed sky spectrum,
which might be obtained by dedicated IFUs, on-off dithers, or using
object-free spatial pixels in the same IFU. Each sky spectrum is then
subtracted from the science spectrum for each spatial element,
producing a sky-subtracted output cube.  This technique is idealised
slightly in the sense that there is little difference between the
simulated sky and that in the science data, which could lead to
residuals in the case of strong variations of the OH lines. However,
OH residuals are present in the simulated spectra due to
differences in the added noise to the science and sky components.

In specifying the detector properties we adopted parameters from tests
of Hawaii-2RG IR arrays by \citet{f06}: a dark current of
0.01e$^-$/pixel/s and a lowest read-out noise of 2.3e$^-$/pixel
\citep[achieved using `Fowler' sampling, see][]{fg90}.  Testing of the four Hawaii-2RG arrays destined
for KMOS has yielded a read-noise of less than 10e$^-$/pixel from
double-correlated sampling \citep{kmos10}; Fowler sampling of the same
arrays should approach the best-case results from \citeauthor{f06}
with existing technology.  Detector development is ongoing in parallel
to plans for the ELTs (e.g. development of 4k\,$\times$\,4k arrays),
such that detector cosmetics/properties will not limit ELT
observations of the type that we describe here.
Long detector integration times (1800s) were adopted, both for
simplicity in the simulations and also to ensure that the observations
at $R$\,$=$\,10000 are background limited (see
Section~\ref{results_10000} for further discussion).

Our primary motivation is to explore the potential of the $J$-band
{\em spectral} diagnostics, rather than a detailed trade-off study of
spatial performance and required AO correction.  In our simulations we
adopt the baseline EAGLE specifications for IFU slice-width
(37.5\,milliarcseconds), informed by the simulated observations of
high-redshift galaxies \citeauthor{mp08} (2008, 2010a). This leads to
a (two spatial-pixel sampled) resolution element of
75\,mas on the sky.  In terms of stellar observations this
can be considered an intermediate level of AO correction, yielding
excellent image quality over a relatively large patrol field for
efficient surveys.  

Note that many of the targets for absorption-line spectroscopy with
the ELTs are already known from deep {\em HST} imaging, but are beyond
the reach of 8-10\,m class facilities. Seeing-limited or Ground-Layer
Adaptive Optics (GLAO) observations will not deliver sufficient
contrast to match such imaging \citep[e.g.][]{c08}, whereas the EAGLE
sampling is commensurate with the {\em HST} pixel scale -- i.e. if a
target is resolved with {\em HST}, then EAGLE is well-matched to
provide follow-up.  The Near-IR Camera (NIRCam) for the 
{\em James Webb Space Telescope (JWST)} also has a
similar pixel scale.

In the densest regions in external galaxies or stellar clusters, finer
angular resolution/greater contrast will be required (thus obtained
over a smaller field).  The spectroscopic methods discussed here will
be equally valid for instruments aimed at probing finer scales such as
HARMONI for the E-ELT \citep{t10} and IRIS for the TMT \citep{larkin},
but detailed performance simulations would require tailored PSFs,
beyond the scope of the current work.

\begin{table}
\begin{center}
\caption{Parameters used in the {\sc websim} simulations.}\label{simspecs}       
\begin{tabular}{ll}
\hline\noalign{\smallskip}
Parameter & Value(s)\\
\noalign{\smallskip}\hline\noalign{\smallskip}
Diameter of primary (M$_{1}$) & 42\,m \\
Effective central obscuration & 9\% \\
Throughput (telescope) & $\ge$80\% \\
Throughput (instrument, inc. detectors) & 30\% @0.850\,$\mu$m \\
& 35\% @1.175\,$\mu$m\\
IFU slice width & 37.5\,mas \\
Spectral resolving power & 4000 \& 10000\\
Dark current & 0.01 e$^-$/pixel/s \\
Read-out noise & 2.3 e$^-$/pixel \\
Exposure time & 20\,$\times$\,1800\,s \\
\noalign{\smallskip}\hline
\end{tabular}
\end{center}
\end{table}

\subsection{Simulated PSFs}\label{moaopsfs}
The EAGLE concept employs multi-object adaptive optics (MOAO) to
provide significantly improved image quality for selected target
fields within the focal plane \citep[first advanced for the FALCON
concept, e.g.][]{agh07}.  The EAGLE baseline design uses an array of
six laser guide stars (at a radius of 3\farcm75) and five natural
guide stars (NGS) to map the atmospheric turbulence \citep{gr10}. The adaptive
mirror in the E-ELT (M$_{4}$) will be used primarily to correct the
ground-layer contribution, while the higher-altitude turbulence will
be corrected by deformable mirrors in each science channel.  The
baseline design of EAGLE has an un-obstructed field-of-view with a
diameter of 5$'$, but the regions between the optics to monitor
the laser guide stars are also accessible, giving a total
field-of-view equivalent to a diameter of 7$'$.

\begin{figure*}
\begin{center}
\includegraphics[scale=1.15]{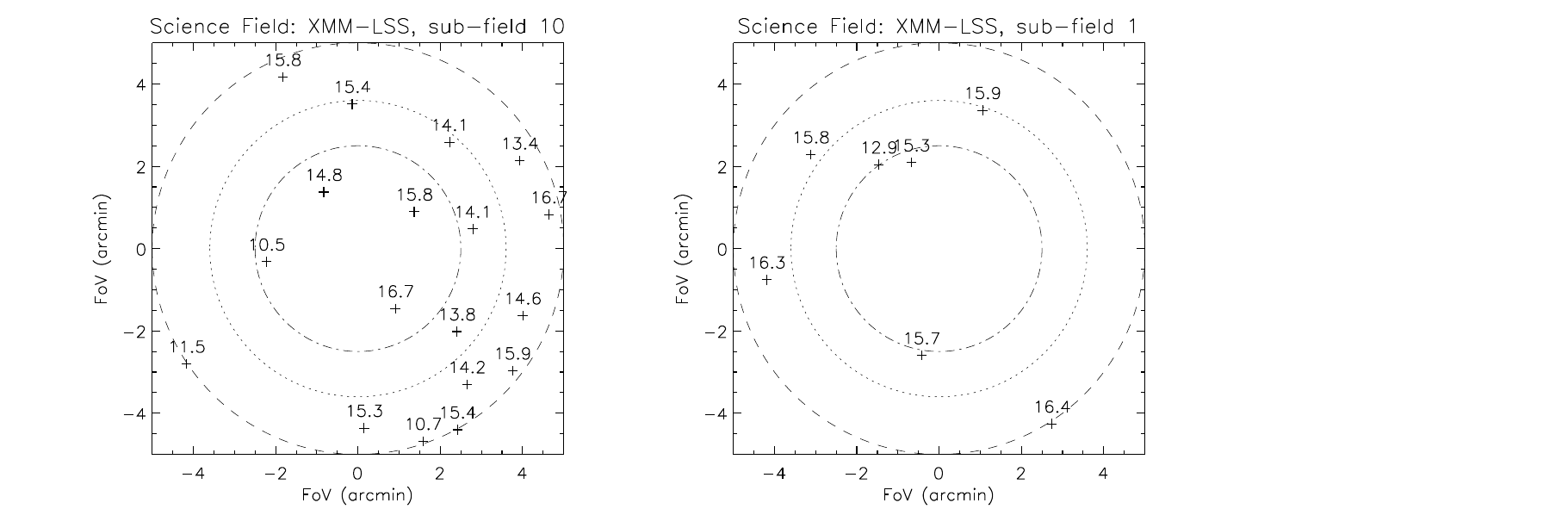}
\caption{Configurations of potential natural guide stars (NGS)
used in the simulated PSFs (labelled by their $R$-band magnitudes).
The overlaid circles have diameters of 5$'$, 7.3$'$, and 10$'$.  {\it
Left:} `good' NGS configuration, with multiple stars available. {\it
Right:} `poor' NGS configuration, with only four stars within the intermediate
circle.}\label{ngs}
\end{center}
\end{figure*}

An extensive set of MOAO simulations were calculated as part of the
EAGLE study \citep{gr10,gr10b}, paying particular attention to the
location and magnitude of potential NGS (with $R$\,$<$\,17) in the
target fields. The PSFs used here were generated
using the ONERA wide-field AO analytical code \citep{n08} with two real NGS
configurations which, given the spatial distribution and magnitude of
the available guide stars, are illustrative of the range of likely
performance.  These were initially selected for high-redshift
simulations, so were taken for two example pointings within the area
of the {\em XMM-Newton} Large-scale Structure Survey \citep{xmm-lss},
as shown in Figure~\ref{ngs}.  Five NGS were selected from those
available in the `good' configuration (left-hand panel of
Figure~\ref{ngs}, which includes one bright target with
$R$\,$=$\,10.5); only four NGS were available in the `poor'
configuration.  As discussed by \citeauthor{e10} (2010b), the
capability to use NGS beyond a 5$'$ diameter leads to excellent sky
coverage.

In practise, the exact AO correction for each sub-field/target is a
function of its location in the patrol field with respect to the NGS.
Calculating tailored PSFs for each potential target is not practical,
so we employ median PSFs (for both `good' and `poor' NGS
configurations) derived from a grid of points across the EAGLE patrol
field from the analytical code.  Typical variations of the ensquared
energy (EE) across the patrol field are of order $\pm$10\% -- a key
part of the planning for such observations will be to optimise the
selection of targets in terms of the expected AO performance.

Assuming a `Paranal-like' set of conditions, we considered two sets of
seeing in the simulations (with the same vertical turbulence profile):
0\farcs65 at \lam\,$=$\,0.5$\mu$m at zenith \citep[the mean VLT
seeing;][]{s08} and 0\farcs90 at a zenith distance (ZD) of 35$^\circ$,
illustrative of the performance that one might expect from execution
of a `Large Programme'-like survey.  The EE in 75\,mas for
each PSF is summarised in Table~\ref{psfs}.

These performances are knowingly optimistic in the respect that
fitting, tomography and noise errors were included, but additional
error terms will contribute to the final performance.  \citet{gr10b}
evaluated the additional wavefront error from these aberrations, which
account for the limitations of the tomographic model, calibration, and
operating in open loop.  The EE degradation can be approximated by a
factor of e$^{- \sigma \phi ^{2}}$, where $\sigma \phi ^{2}$ is the
phase variance (in radian$^2$) of the wavefront error at the observed
wavelength.  This leads to a scaling of the EE in the simulated PSFs
by a factor of 0.69 in $H$ (calculated at 1.65\,$\mu$m), 0.52 in $J$
(1.25\,$\mu$m), and 0.25 in $I$ (0.85\,$\mu$m). Although the
morphology of the PSF will also change slightly, in calculation of the
resulting signal-to-noise (S/N) ratios in the simulated spectra (next section)
and interpretation of our results (Section~\ref{discussion})
we consider only the more important equivalent loss in sensitivity. As
noted before, the current study is primarily intended as a
proof-of-concept study for the $J$-band diagnostics -- the exact AO
performances will continue to evolve with the telescope and instrument
design, and experience from on-sky demonstrators such as CANARY on the
William Herschel Telescope \citep[e.g.][]{gmh10}.

\begin{table}
\begin{center}
\caption{Ensquared energy (EE) in 75\,milliarcseconds in the simulated MOAO PSFs used
in the $I$- and $J$-band simulations presented here. Note that these
do not include some of the potential error terms (see Section~\ref{moaopsfs}), and are 
scaled accordingly in the discussion.}\label{psfs}       
\begin{tabular}{cclllc}
\hline\noalign{\smallskip}
Band & \lam$_{\rm c}$ [$\mu$m] & Seeing & ZD & NGS & EE [75\,mas] \\
\noalign{\smallskip}\hline\noalign{\smallskip}
$I$ & 0.85 & 0\farcs65 & $\phantom{3}$0$^\circ$ & Good & 43\% \\
$I$ & 0.85 & 0\farcs65 & $\phantom{3}$0$^\circ$ & Poor & 37\% \\
$I$ & 0.85 & 0\farcs90 & 35$^\circ$ & Good & 25\% \\
$I$ & 0.85 & 0\farcs90 & 35$^\circ$& Poor & 19\% \\
\noalign{\smallskip}\hline
$J$ & 1.25 & 0\farcs65 & $\phantom{3}$0$^\circ$ & Good & 64\% \\
$J$ & 1.25 & 0\farcs65 & $\phantom{3}$0$^\circ$ & Poor & 58\% \\
$J$ & 1.25 & 0\farcs90 & 35$^\circ$ & Good & 48\% \\
$J$ & 1.25 & 0\farcs90 & 35$^\circ$& Poor & 40\% \\
\noalign{\smallskip}\hline
\end{tabular}
\vspace{-1cm}
\end{center}
\end{table}

\subsection{Template spectra}

Our simulations adopt two spectral templates from the grid of
low-temperature stellar atmospheres calculated with the {\sc marcs}
code \citep{g08}. These have an effective temperature (\Teff) of 3800\,K
(corresponding to an equivalent spectral type of approximately M0),
\logg\,$=$\,0.0, and a microturbulence ($\xi$) of 2\,km\,s$^{-1}$.
Models for two metallicities ($Z$) were used, with \logz\,$=$\,0.0 and
$-$1.0 (normalised to Solar).  The latter model (i.e. 10\% Solar)
serves as a test of our methods in the low metallicity regime,
comparable to, e.g., results for some of the young supergiants
observed in the WLM dwarf irregular galaxy \citep{u08}.
Further details of the {\sc marcs} models, in the context of the
current study, are given by DKF10.

At Solar metallicity we also calculated a set of $I$-band simulations
to inform discussion of the $J$-band results in
Section~\ref{discussion}.  The $I$-band includes the calcium triplet
(CaT, spanning 0.85-0.87$\mu$m), which has become an increasingly
ubiquitous diagnostic of stellar metallicities and radial velocities
in nearby galaxies \citep[e.g.][]{t04}.  The CaT is an empirical
calibration of metallicity (which employs the equivalent width of the
Ca~\2 lines) rather than a direct diagnostic.  However, careful
calibration yields a robust relationship over a large range of
metallicity \citep [{$-$2.5\,$<$\,[Fe/H]\,$<$\,$-$0.2};][]{c04}.  Our
primary interest in the CaT is simply to obtain an estimate of the
continuum sensitivities, so only one template (\logz\,$=$\,0.0) is
required.

The templates were convolved by the relevant spectral resolving power,
and then binned such that each resolution element is sampled by two
pixels.  The templates were restricted to spectral windows of
\lam\lam1.15-1.22\,$\mu$m ($J$-band) and \lam\lam0.840-0.875\,$\mu$m
($I$-band), with artificial continuum regions appended (with an
intensity set to unity) to provide straightforward estimates of the
S/N in the simulated spectra.

\begin{table*}
\begin{center}
\caption{Summary of continuum signal-to-noise (S/N) obtained per two-pixel resolution element for simulated $J$-band
spectroscopy ($t_{\rm exp}$\,$=$\,10\,hrs, $R$\,$=$\,4000 \&
10000). The quoted magnitudes are after scaling by the estimated
uncertainties in the simulated PSFs (see Section~\ref{moaopsfs} for
further discussion and for details on the configurations of natural guide stars, `NGS').}
\label{Jsims}
\begin{tabular}{cccccccccc}
\hline\noalign{\smallskip}
& & \multicolumn{4}{c}{$R$\,$=$\,4000} &
\multicolumn{4}{c}{$R$\,$=$\,10000} \\ & &
\multicolumn{2}{c}{Seeing\,$=$\,0{\mbox{\ensuremath{.\!\!^{\prime\prime}}}}9
@ ZD=35$^\circ$} &
\multicolumn{2}{c}{Seeing\,$=$\,0{\mbox{\ensuremath{.\!\!^{\prime\prime}}}}65 @ ZD=0$^\circ$} &
\multicolumn{2}{c}{Seeing\,$=$\,0{\mbox{\ensuremath{.\!\!^{\prime\prime}}}}9 @ ZD=35$^\circ$} &
\multicolumn{2}{c}{Seeing\,$=$\,0{\mbox{\ensuremath{.\!\!^{\prime\prime}}}}65 @ ZD=0$^\circ$} \\
$J_{\rm VEGA}$ & $Z$ & NGS good & NGS poor & NGS good & NGS poor &
NGS good & NGS poor & NGS good & NGS poor \\
\noalign{\smallskip}\hline\noalign{\smallskip}
20.75 & \pmin0.0 & 266\,$\pm$\,22 & 219\,$\pm$\,16 & 306\,$\pm$\,15 & 307\,$\pm$\,40 & 163\,$\pm$\,12 & 138\,$\pm$\,6 & 192\,$\pm$\,10 & 177\,$\pm$\,\pA7 \\	   
        &    $-$1.0 & 260\,$\pm$\,23 & 231\,$\pm$\,29 & 329\,$\pm$\,24 & 308\,$\pm$\,28 & 169\,$\pm$\,\pA8 & 140\,$\pm$\,9 & 200\,$\pm$\,18 & 178\,$\pm$\,\pA9 \\	   
\noalign{\smallskip}
21.25 & \pmin0.0 & 209\,$\pm$\,26 & 173\,$\pm$\,20 & 261\,$\pm$\,24 & 221\,$\pm$\,26 & 126\,$\pm$\,13 & 100\,$\pm$\,6 & 148\,$\pm$\,10 & 136\,$\pm$\,\pA9 \\  
        &    $-$1.0 & 203\,$\pm$\,18 & 170\,$\pm$\,19 & 234\,$\pm$\,21 & 227\,$\pm$\,23 & 124\,$\pm$\,11 & 101\,$\pm$\,6 & 155\,$\pm$\,11 & 134\,$\pm$\,12 \\  
\noalign{\smallskip}
21.75 & \pmin0.0 & 143\,$\pm$\,16 & 122\,$\pm$\,20 & 180\,$\pm$\,14 & 167\,$\pm$\,16 & \pA92\,$\pm$\,\pA7 & \pA74\,$\pm$\,5 & 110\,$\pm$\,10 & 102\,$\pm$\,\pA3 \\
    &    $-$1.0 & 140\,$\pm$\,11 & 122\,$\pm$\,\pA7 & 184\,$\pm$\,23 & 160\,$\pm$\,14 & \pA89\,$\pm$\,\pA6 & \pA75\,$\pm$\,6 & 111\,$\pm$\,\pA7 & 101\,$\pm$\,\pA7 \\
\noalign{\smallskip}
22.25 & \pmin0.0 & 106\,$\pm$\,\pA9 & \pA84\,$\pm$\,\pA6 & 137\,$\pm$\,14 & 127\,$\pm$\,24 & \pA64\,$\pm$\,\pA6 & \pA50\,$\pm$\,2 & \pA80\,$\pm$\,\pA7 & \pA69\,$\pm$\,\pA5 \\
  &    $-$1.0 & \pA99\,$\pm$\,10 & \pA81\,$\pm$\,\pA5 & 129\,$\pm$\,20 & 114\,$\pm$\,10 & \pA62\,$\pm$\,\pA4 & \pA50\,$\pm$\,5 & \pA78\,$\pm$\,\pA6 & \pA72\,$\pm$\,\pA6 \\
\noalign{\smallskip}
22.75 & \pmin0.0 & \pA71\,$\pm$\,\pA7 & \pA60\,$\pm$\,\pA6 & \pA98\,$\pm$\,10 & \pA84\,$\pm$\,\pA4 & \pA45\,$\pm$\,\pA3 & \pA35\,$\pm$\,2 & \pA55\,$\pm$\,\pA4 & \pA53\,$\pm$\,\pA5 \\
        &    $-$1.0 & \pA72\,$\pm$\,11 & \pA58\,$\pm$\,\pA8 & \pA96\,$\pm$\,14 & \pA79\,$\pm$\,\pA7 & \pA44\,$\pm$\,\pA3 & \pA35\,$\pm$\,2 & \pA55\,$\pm$\,\pA4 & \pA49\,$\pm$\,\pA2 \\
\noalign{\smallskip}
23.25 & \pmin0.0 & \pA54\,$\pm$\,\pA5 & \pA40\,$\pm$\,\pA5 & \pA65\,$\pm$\,\pA8 & \pA62\,$\pm$\pA\,6 & \pA30\,$\pm$\,\pA2 & \pA24\,$\pm$\,2 & \pA38\,$\pm$\,\pA3 & \pA35\,$\pm$\,\pA2 \\
        &    $-$1.0 & \pA50\,$\pm$\,\pA8 & \pA39\,$\pm$\,\pA7 & \pA62\,$\pm$\,\pA6 & \pA56\,$\pm$\,10 & \pA30\,$\pm$\,\pA2 & \pA23\,$\pm$\,2 & \pA38\,$\pm$\,\pA2 & \pA36\,$\pm$\,\pA2 \\
\noalign{\smallskip}
23.75 & \pmin 0.0 & \pA35\,$\pm$\,\pA4 & \pA27\,$\pm$\,\pA3 & \pA45\,$\pm$\,\pA6 & \pA40\,$\pm$\,\pA3 & $-$ & $-$ & $-$ & $-$ \\
        &     $-$1.0 & \pA32\,$\pm$\,\pA5 & \pA27\,$\pm$\,\pA4 & \pA40\,$\pm$\,\pA4 & \pA39\,$\pm$\,\pA5 & $-$ & $-$ & $-$ & $-$ \\
\noalign{\smallskip}\hline
\end{tabular}
\end{center}
\end{table*}

The {\sc websim} routines operate on fluxes ($F$) in physical units
(either ergs/s/cm$^2$/Hz or ergs/s/cm$^2$/\AA), so the input model
templates were scaled to the desired observed magnitudes ($M$, in the
Vega system), using:
\begin{center}
\begin{equation}
F_{\nu} = \frac{F_{M_{\nu}=0}}{10^{\,(M_{\nu}/2.5)}}~,
\end{equation}
\end{center}

where $F_{(I =0)}$\,$=$\,2550\,Jy and $F_{(J=0)}$\,$=$\,1600\,Jy
\citep[][respectively]{b79,crl85}\footnote{The passband of {Cousins $I$} is somewhat different to that of
the F814W {\em HST} filters, but the typical magnitude
offset is small: $I$\,$-$\,F814W $\le$\,0.05$^{\rm m}$ \citep{h95}.}.

The simulated IFU observations are of an isolated point source centred
in the central (37.5\,$\times$\,37.5 mas) spatial pixel, which is then
extracted to yield the object spectrum for each target magnitude/set
of conditions.  The image quality of the simulated PSFs is
sufficiently good that, in this idealised case of a centrally-located
point source, extraction and merging with adjacent spatial elements
does not improve the S/N of the final spectrum; in real EAGLE
observations, more sophisticated extraction routines (e.g.,
PSF-fitting) would likely be required.

\subsection{Simulation results}

\begin{figure*}
\begin{center}
\includegraphics[scale=0.45]{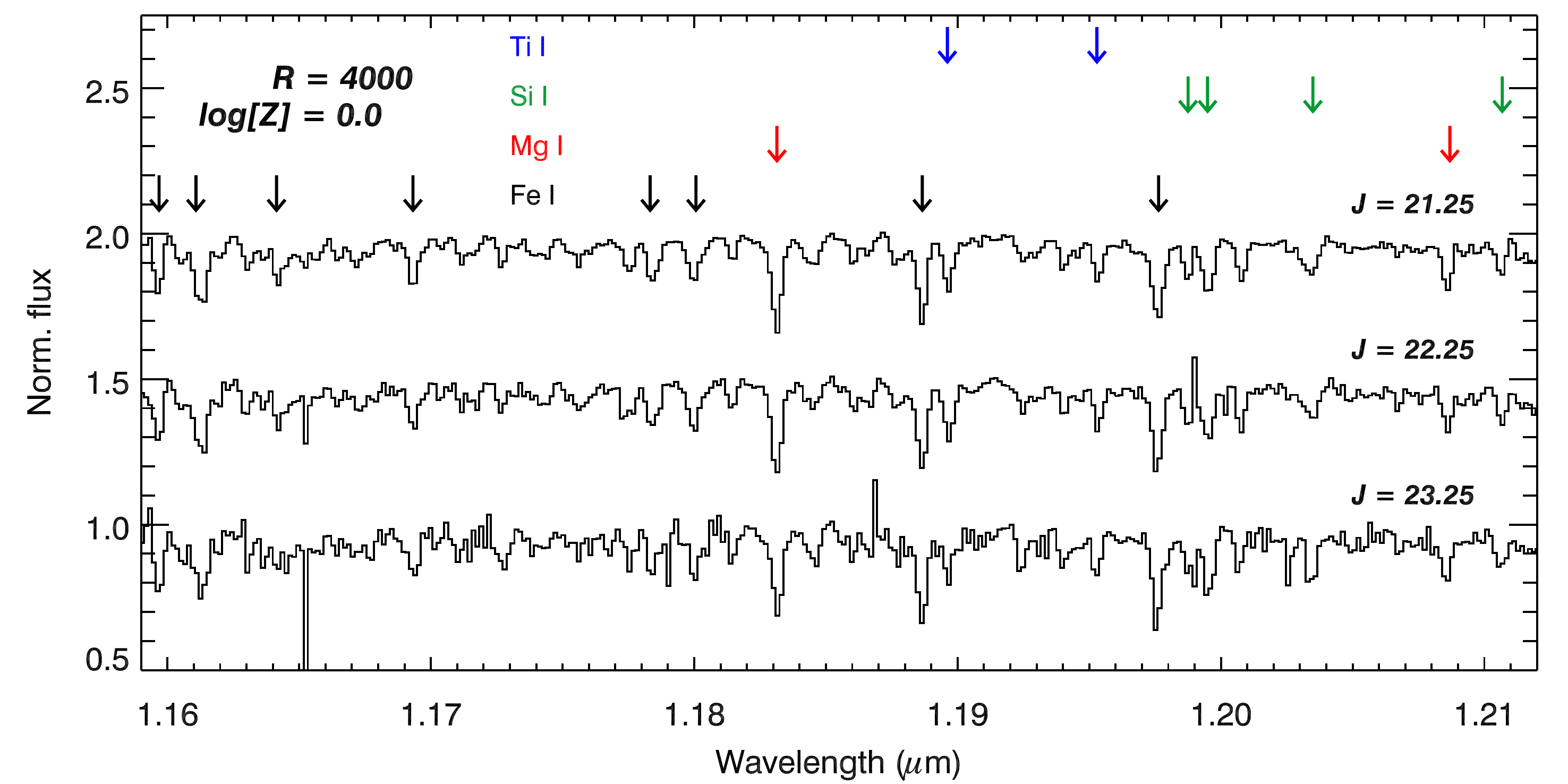}
\vspace*{0.2in}\\
\includegraphics[scale=0.45]{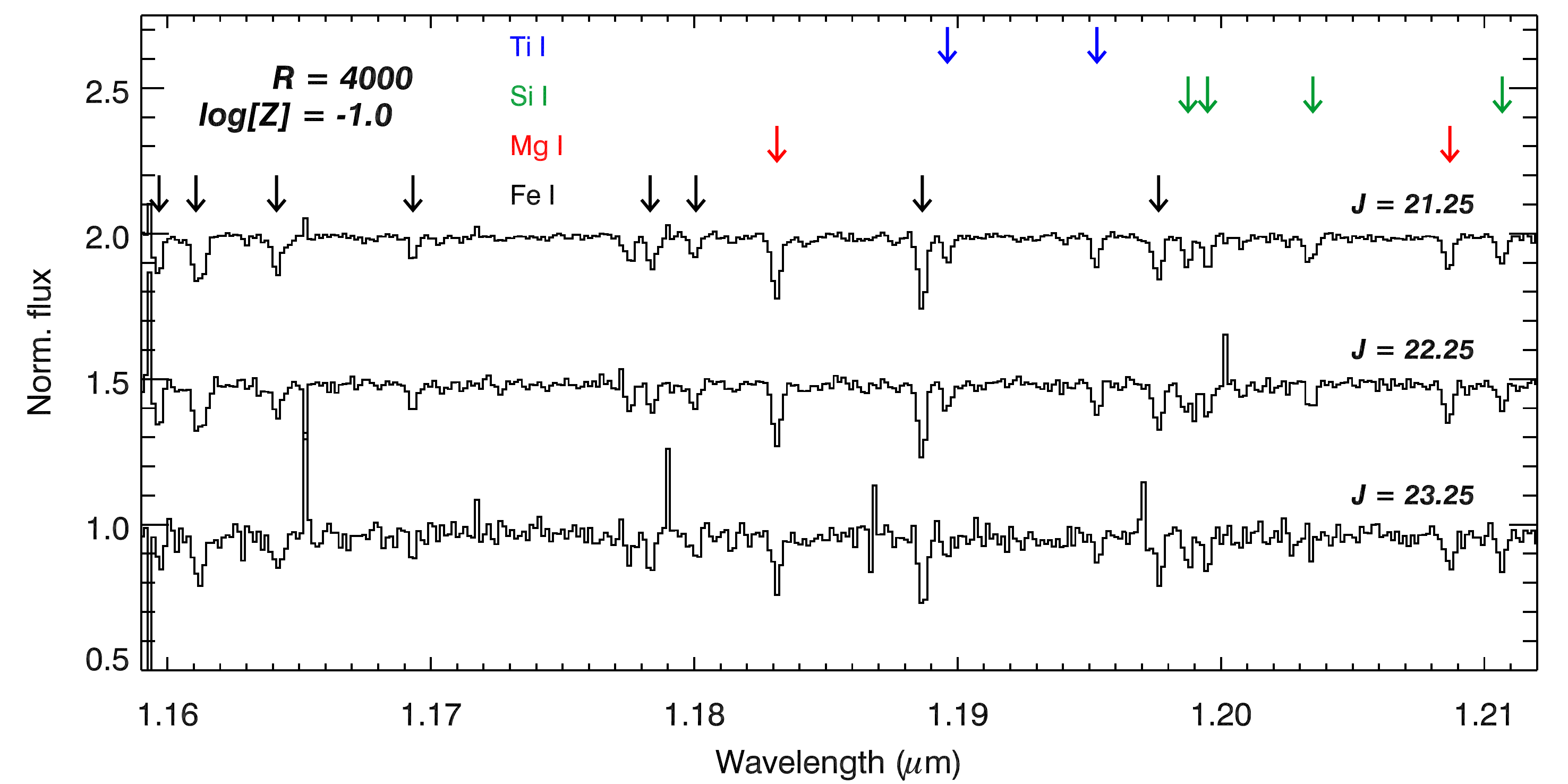}
\caption{Examples of simulated $R$\,$=$\,4000 spectra for $J$\,$=$\,21.25, 22.25, and 23.25
(with seeing of 0\farcs9, ZD\,$=$\,35$^\circ$, and the `good' NGS
configuration). Upper panel: Solar metallicity simulations
(\logz\,$=$\,0.0); lower panel: metal-poor simulations
(\logz\,$=$\,$-$1.0). Identified lines, from left-to-right by species,
are: Ti~\1 \lam\lam1.1896, 1.1953\,$\mu$m; Si~\1 \lam\lam 1.1988,
1.1995, 1.2035, 1.2107\,$\mu$m; Mg~\1 \lam\lam1.1831, 1.2087\,$\mu$m;
Fe~\1 \lam\lam1.1597, 1.1611, 1.1641, 1.1693, 1.1783, 1.1801, 1.1887,
1.1976\,$\mu$m. Residuals from the sky OH lines can be seen in places, e.g. the 
emission `spikes' in the simulated metal-poor spectrum for $J$\,$=$\,23.25.}\label{simspec_J4000}
\end{center}
\end{figure*}

\begin{figure*}
\begin{center}
\includegraphics[scale=0.45]{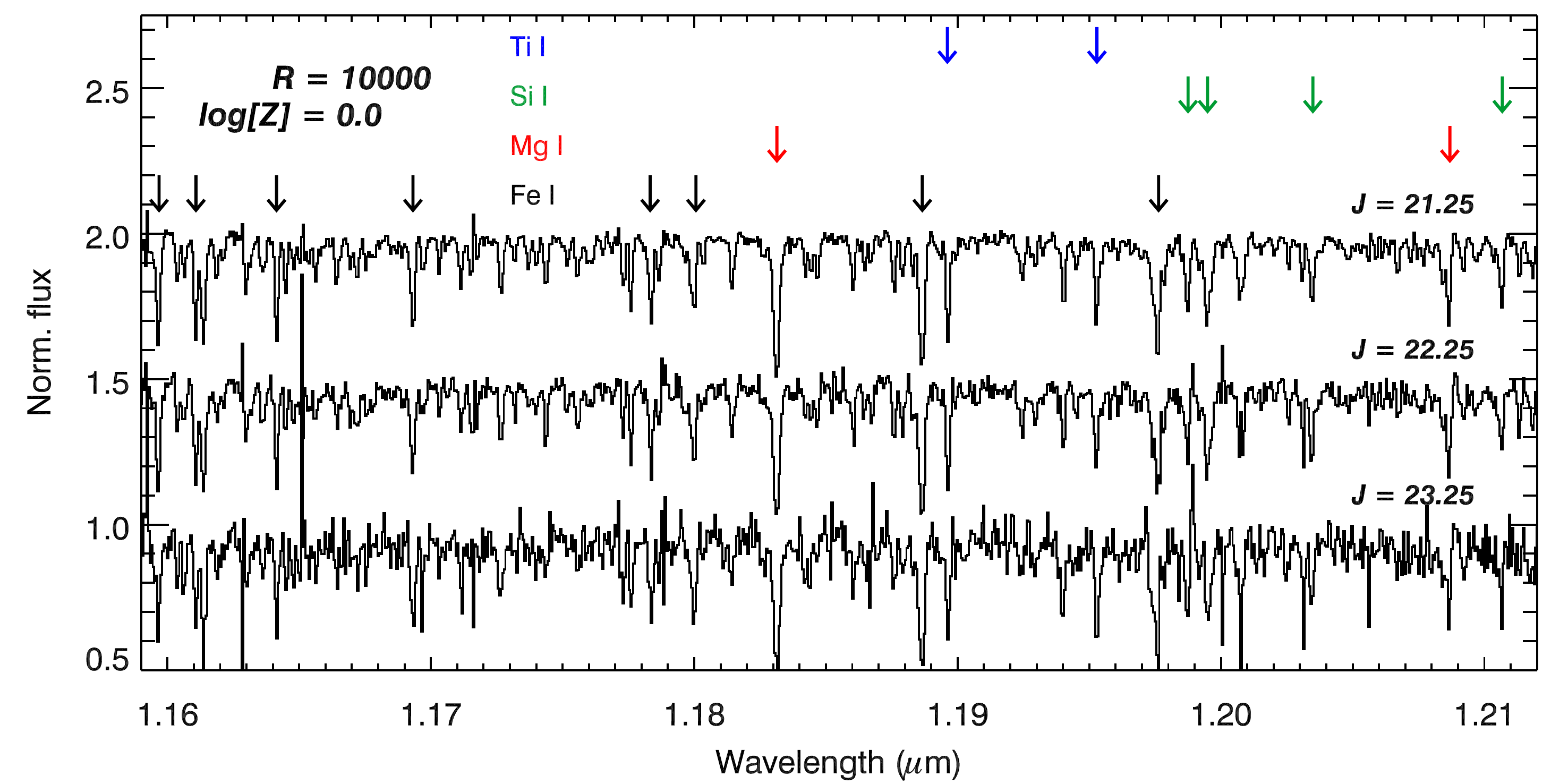}
\vspace*{0.2in}\\
\includegraphics[scale=0.45]{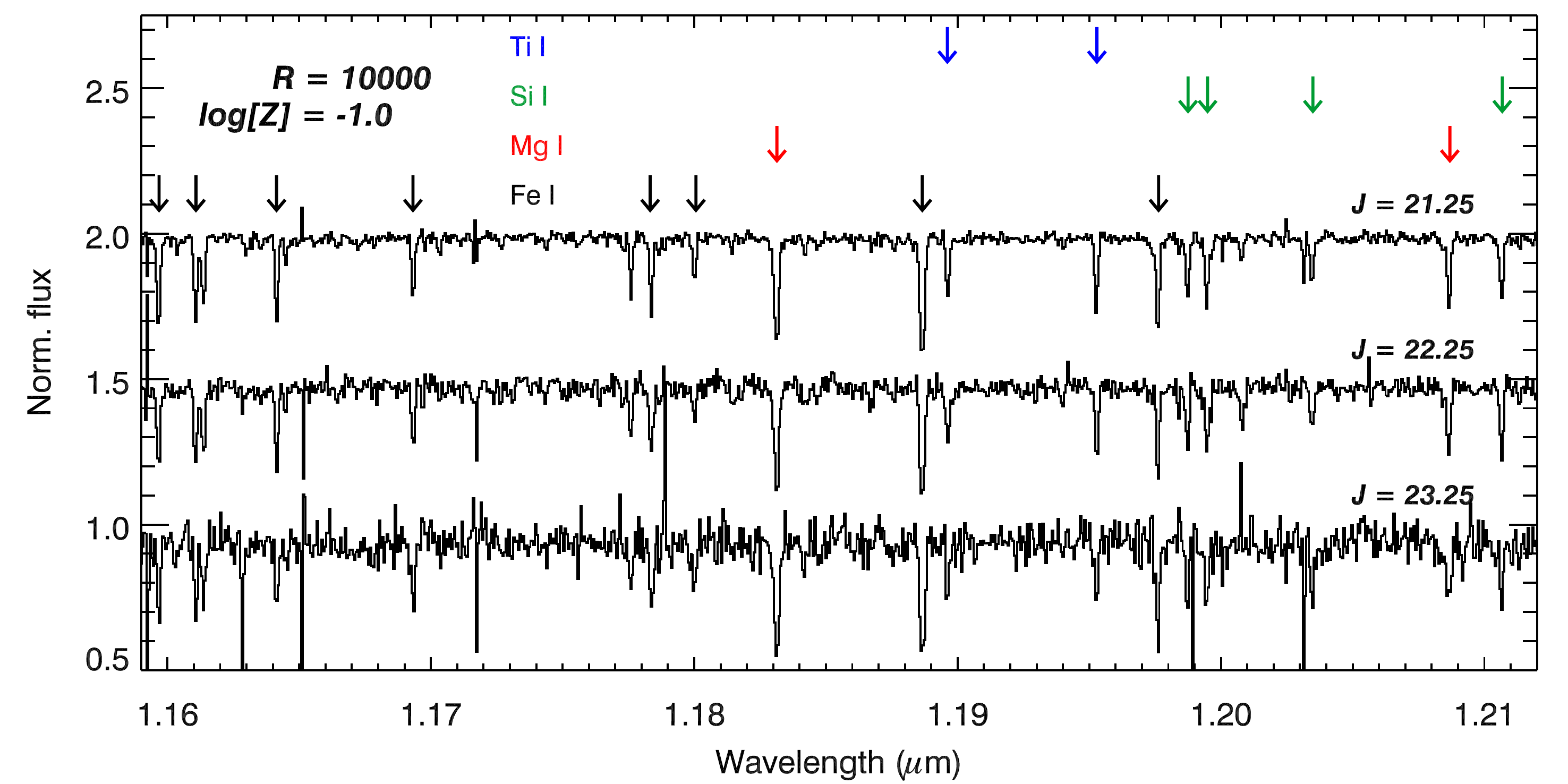}
\caption{Examples of simulated $R$\,$=$\,10000 spectra for $J$\,$=$\,21.25, 22.25, and 23.25
(with seeing of 0\farcs9, ZD\,$=$\,35$^\circ$, and the `good' NGS configuration). Upper panel:
Solar metallicity simulations (\logz\,$=$\,0.0); lower panel: metal-poor 
simulations (\logz\,$=$\,$-$1.0). Identified lines are the same as those in Figure~\ref{simspec_J4000}.}\label{simspec_J10000}
\includegraphics[scale=0.45]{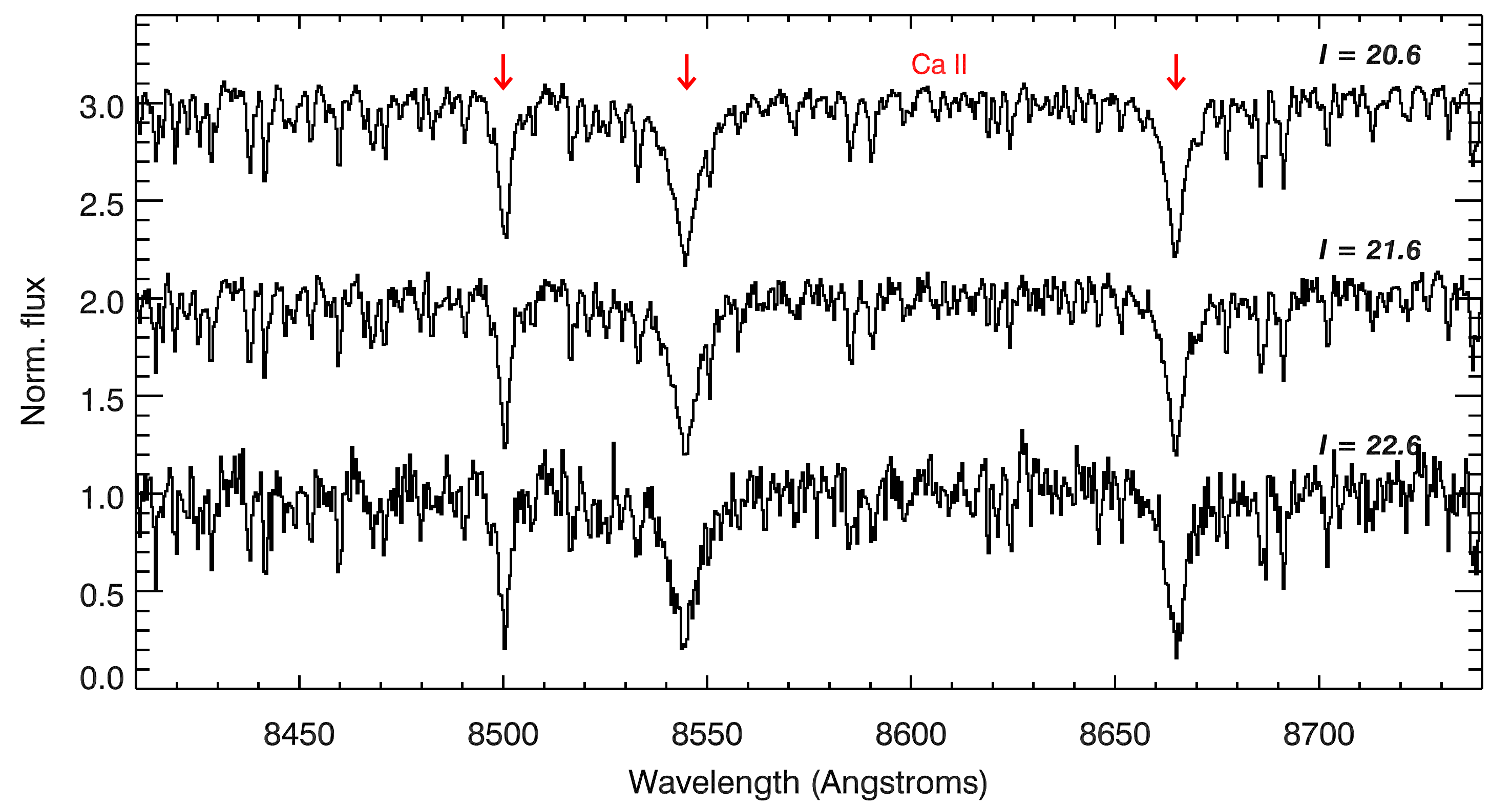}
\caption{Examples of simulated CaT spectra ($R$\,$=$\,10000, \logz\,$=$\,0.0) for $I$\,$=$\,20.6, 21.6, and 22.6,
(with seeing of 0\farcs9 and the `good' NGS configuration). The three Ca~\2 lines identified are \lam\lam8498, 8542, 8662\,\AA.
}\label{simspec_I10000}
\end{center}
\end{figure*}

Each set of simulations was run ten times to provide a meaningful
estimate of the dispersions in the derived metallicities and 
continuum S/N.  Example simulated $J$-band spectra are shown in
Figures~\ref{simspec_J4000} and \ref{simspec_J10000}, with notable
diagnostic lines of Mg~\1, Si~\1, Ti~\1 and Fe~\1 labelled.  Example
$I$-band spectra are shown in Figure~\ref{simspec_I10000}.

\begin{figure}
\begin{center}
\includegraphics[width=8cm]{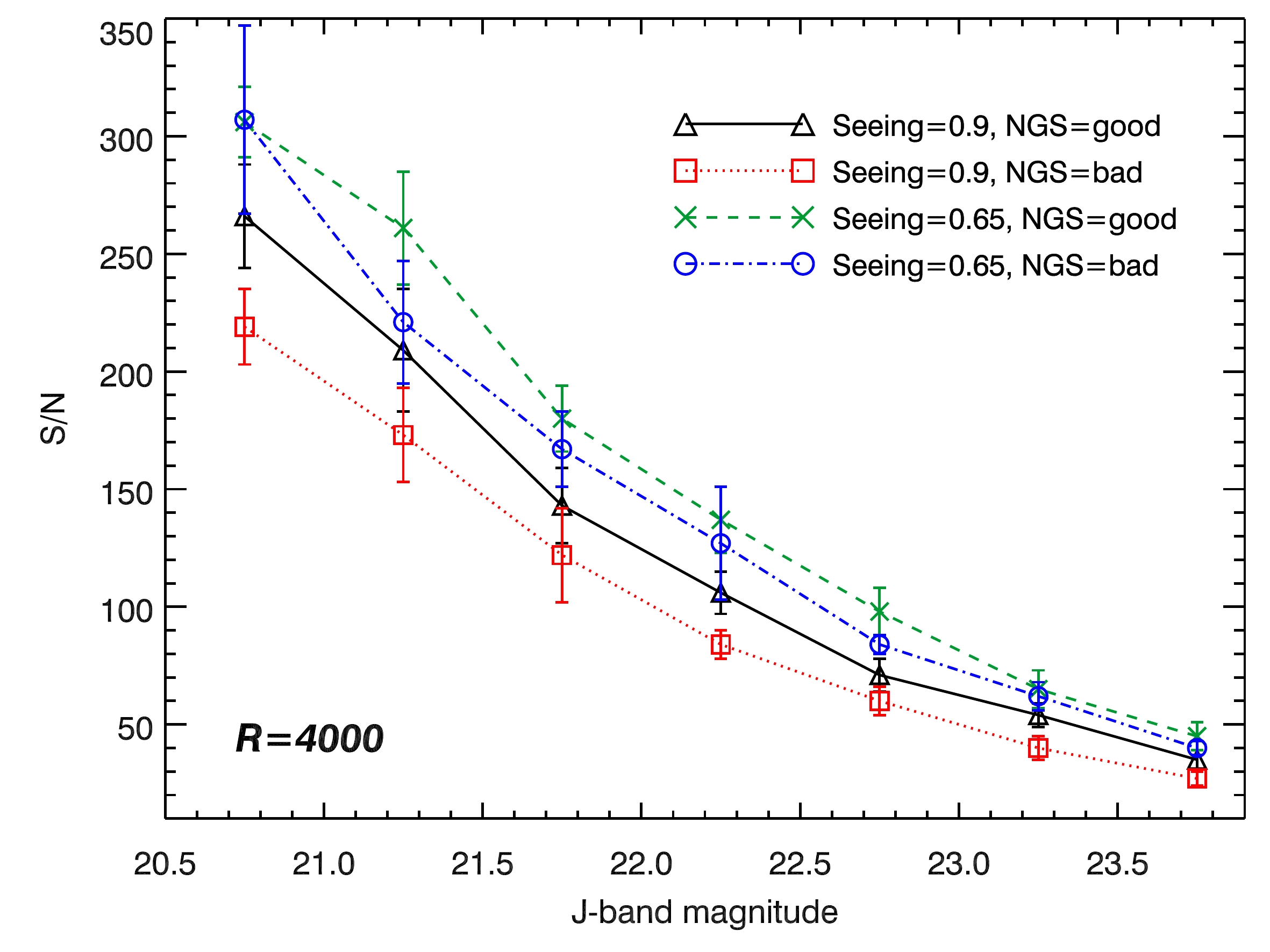}
\caption{Continuum signal-to-noise (S/N) results for simulated $J$-band
spectroscopy ($t_{\rm exp}$\,$=$\,10\,hrs, $R$\,$=$\,4000) from Table~\ref{Jsims}.}\label{SN_4000}
\vspace{0.5cm}
\includegraphics[width=8cm]{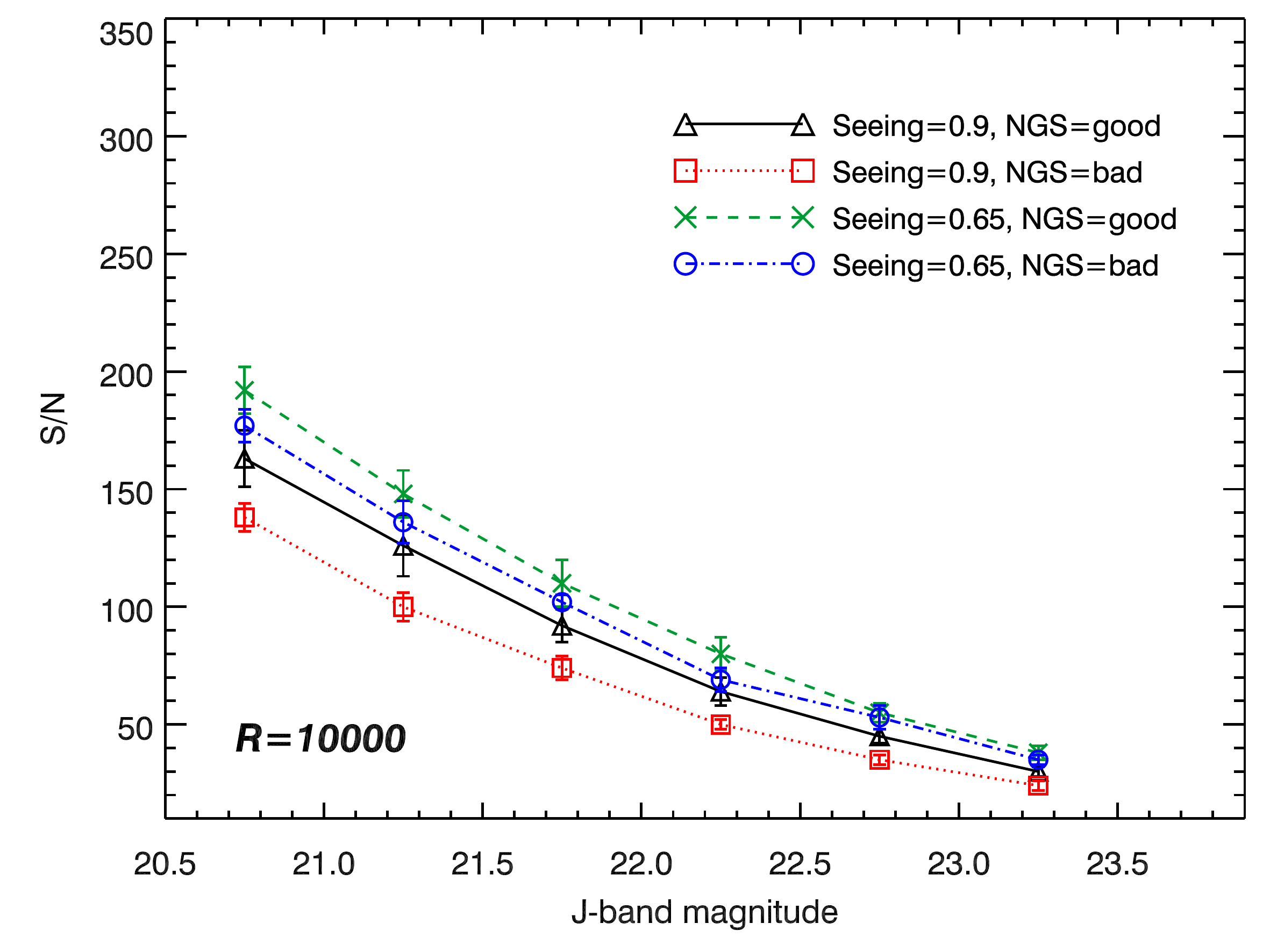}
\caption{Continuum signal-to-noise (S/N) results for simulated $J$-band
spectroscopy ($t_{\rm exp}$\,$=$\,10\,hrs, $R$\,$=$\,10000) from Table~\ref{Jsims}.}\label{SN_10000}
\vspace{0.5cm}
\includegraphics[width=8cm]{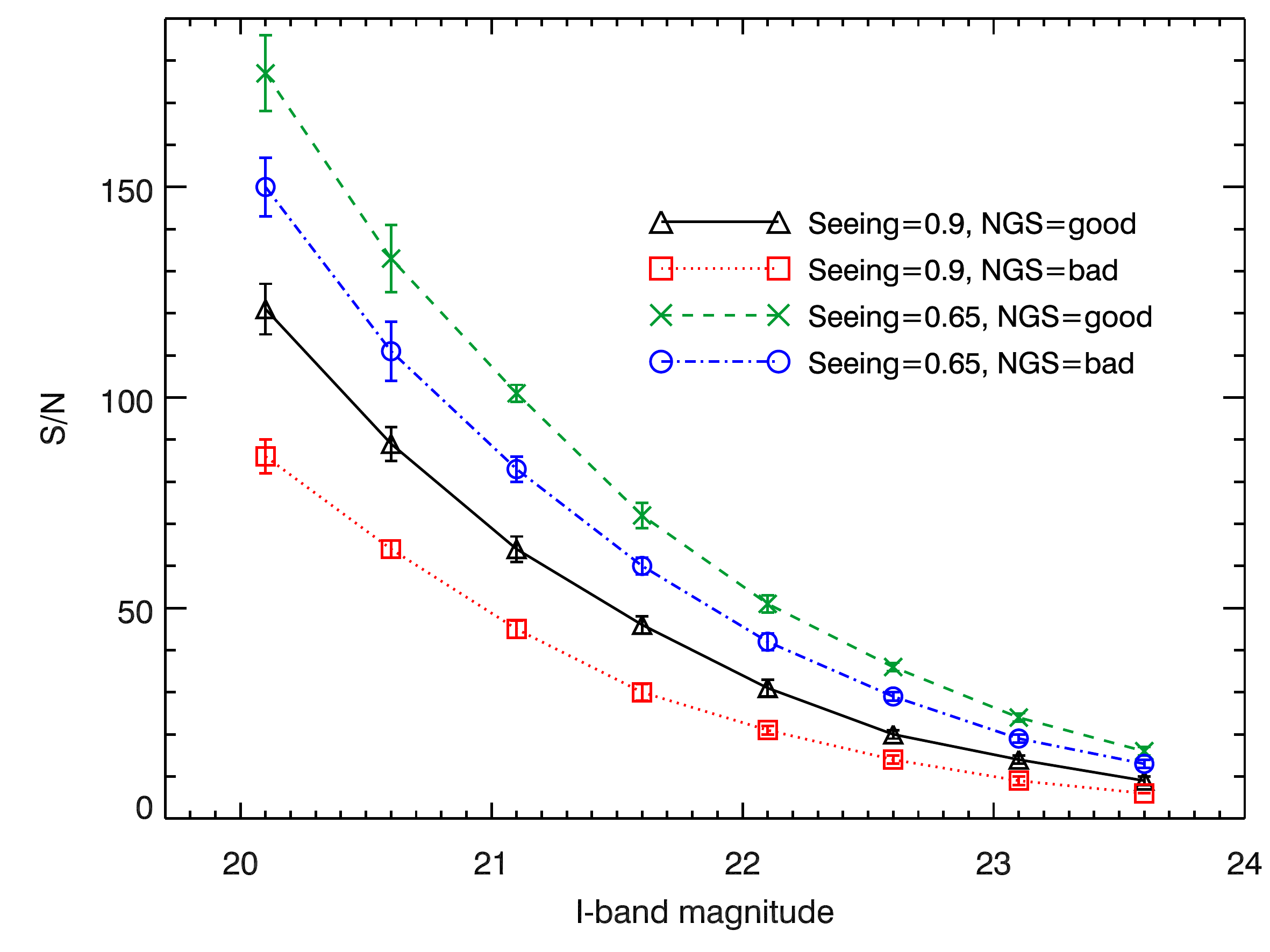}
\caption{Continuum signal-to-noise (S/N) results for simulated $I$-band
spectroscopy ($t_{\rm exp}$\,$=$\,10\,hrs, $R$\,$=$\,10000) from Table~\ref{Isims}.}\label{SN_I}
\end{center}
\end{figure}

The means and standard deviations of the continuum S/N from the $J$-
and $I$-band runs are summarised in Tables~\ref{Jsims} and
\ref{Isims}, respectively; the results are also shown in
Figures~\ref{SN_4000}, \ref{SN_10000}, and \ref{SN_I}.  Following the
discussion in Section~\ref{moaopsfs} regarding the known limitations
of the AO simulations, the quoted magnitudes take into account the
factor of two in intensity in the $J$-band (equivalent to $\Delta
J$\,$=$\,$-$0.75) and the factor of four in the $I$-band (equivalent
to $\Delta I$\,$=$\,$-$1.5), i.e. the effective magnitude of each
simulation is brighter than the input magnitude.  All quoted S/N
results are for spectra extracted from the central spatial pixel of
each cube, per two spectral pixels (set to be the sampling of the
resolution element in each \lam\/ and $R$ combination).

\begin{table}
\begin{center}
\caption{Summary of continuum signal-to-noise (S/N) obtained per two-pixel resolution element
for simulated $I$-band spectroscopy ($t_{\rm exp}$\,$=$\,10\,hrs, $R$\,$=$\,10000).
As in Table~\ref{Jsims}, the quoted magnitudes are after scaling by the estimated
uncertainty in the simulated PSFs (see Section~\ref{moaopsfs}).}\label{Isims}       
\begin{tabular}{ccccc}
\hline\noalign{\smallskip}
& \multicolumn{2}{c}{Seeing\,$=$\,0{\mbox{\ensuremath{.\!\!^{\prime\prime}}}}9 @ ZD=35$^\circ$} &
\multicolumn{2}{c}{Seeing\,$=$\,0{\mbox{\ensuremath{.\!\!^{\prime\prime}}}}65 @ ZD=0$^\circ$} \\
$I_{\rm VEGA}$ & NGS good & NGS poor & NGS good & NGS poor \\
\noalign{\smallskip}\hline\noalign{\smallskip}
20.1 & 121\,$\pm$\,6 & 86\,$\pm$\,4 & 177\,$\pm$\,9 & 150\,$\pm$\,7 \\
20.6 & \pA89\,$\pm$\,4 & 64\,$\pm$\,2 & 133\,$\pm$\,8 & 111\,$\pm$\,7 \\
21.1 & \pA64\,$\pm$\,3 & 45\,$\pm$\,2 & 101\,$\pm$\,2 & \pA83\,$\pm$\,3 \\
21.6 & \pA46\,$\pm$\,2 & 30\,$\pm$\,2 & \pA72\,$\pm$\,3 & \pA60\,$\pm$\,2 \\
22.1 & \pA31\,$\pm$\,2 & 21\,$\pm$\,1 & \pA51\,$\pm$\,2 & \pA42\,$\pm$\,2 \\
22.6 & \pA20\,$\pm$\,1 & 14\,$\pm$\,1 & \pA36\,$\pm$\,1 & \pA29\,$\pm$\,1 \\
23.1 & \pA14\,$\pm$\,1 & \pA9\,$\pm$\,1 & \pA24\,$\pm$\,1 & \pA19\,$\pm$\,1 \\
23.6 & \pA\pA9\,$\pm$\,1 & \pA6\,$\pm$\,0 & \pA16\,$\pm$\,1 & \pA13\,$\pm$\,1 \\
\noalign{\smallskip}\hline
\end{tabular}
\end{center}
\end{table}

\section{Analysis}\label{analysis}

\subsection{$J$-band simulations}\label{results_J}

Our goal in the first part of the analysis is to determine physical
properties from the simulated observations and compare them with those
of the input model, i.e., how successfully can we recover accurate
metallicities at both \logz\,$=$\,0.0 and $-$1.0, and what S/N is
required?

We follow similar methods to those described by DKF10.
Each simulated spectrum is compared with template spectra from the
{\sc marcs} grid, the \chisq\/ is calculated, and the location in the
grid of the \chisq\/ minimum is found. The parameter space searched
includes 3200\,$<$\,\Teff\,$<$\,4000\,K (in steps of 200\,K),
$-$1.5\,$<$\,\logz\,$<$\,$+$1.0 (in steps of 0.25\,dex)\footnote{The
grid does not contain models for \logz\,=\,$-$1.25, so spectra for this
metallicity are interpolated from neighbouring models.},
2\,$<$\,$\xi$\,$<$\,5\,\kms (in steps of 1\,\kms)\footnote{The spectra
at $\xi$\,$=$\,3 and 4\,\kms\/ are interpolated (see DKF10).}, and
\logg\,$=$\,0.0 and $+$1.0.

Two refinements to the methods from DKF10 were necessary for the
analysis presented here.  This concerns the location of the continuum
level in the simulated spectra and the interpolation between model
grid points to refine the metallicity determinations, both of which
are now discussed.

\subsubsection{Continuum placement}

To obtain accurate metallicities from the simulated spectra it is
crucial that the continuum level is determined
correctly. Placing the continuum at the wrong level results in an
incorrect measurement of the normalised line strengths, thus an
erroneous metallicity. DKF10 found the continuum level of their input
spectra by ranking the spectral pixels in order of their (normalised)
flux, and then finding the median value of those pixels with the
greatest flux. In a spectrum comprised of continuum and absorption
lines, the pixels with the greatest flux will have a value equivalent
to the continuum level plus the noise. At high S/N this is a very good
approximation of the continuum level, but at low S/N this can lead to
the estimated continuum level being too high, resulting in
a derived metallicity that is systematically above that of the
input spectrum.

\begin{figure}
\begin{center}
\includegraphics[width=8cm]{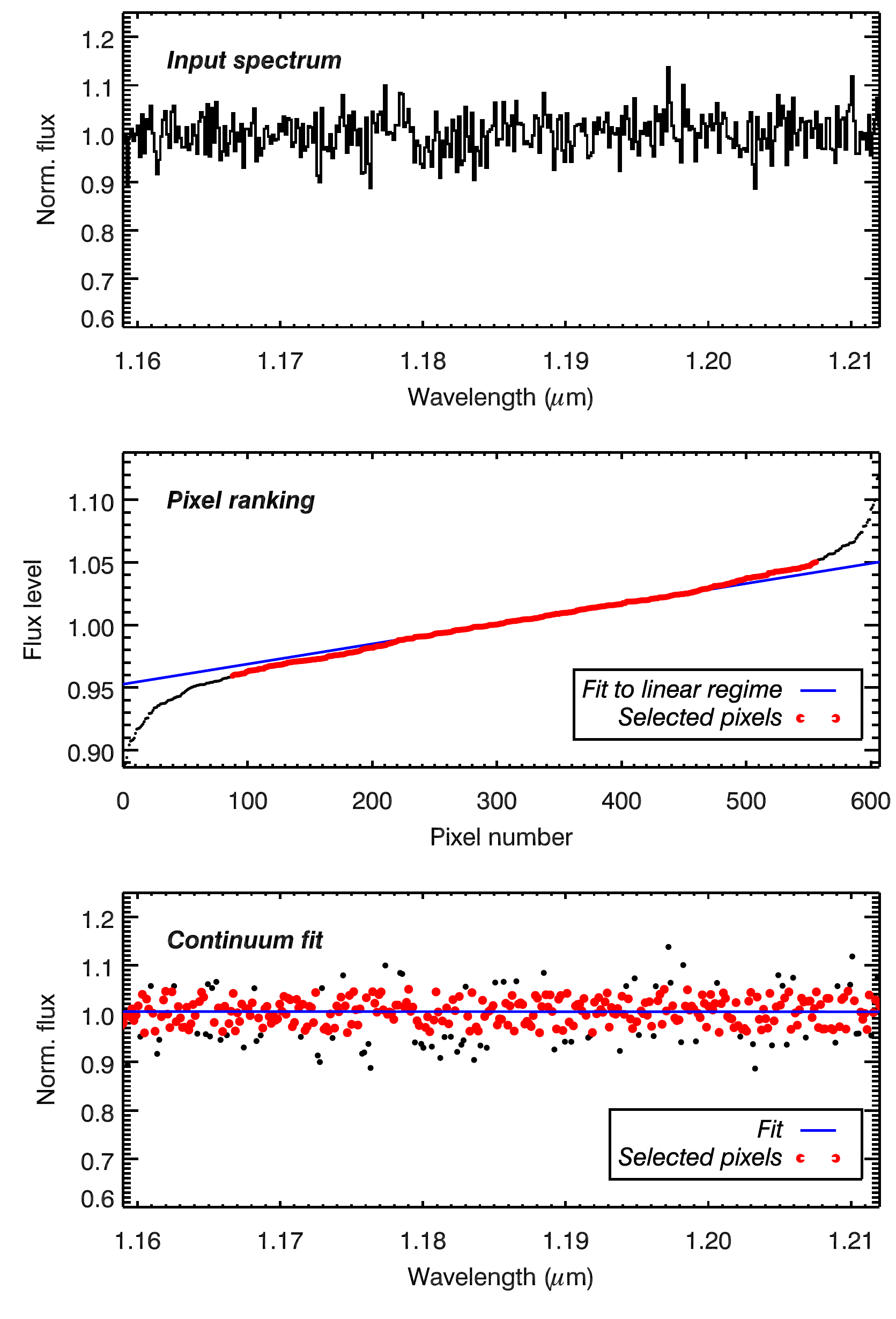}
\caption{Illustration of the continuum-fitting procedure for a
spectrum of pure Gaussian noise. {\it Top panel:} input spectrum with
S/N\,$=$\,50. {\it Centre:} input spectrum plotted in order of
increasing pixel flux (the `pixel ranking' spectrum). The blue
line is a fit to the linear regime, with pixels selected as part of
that regime ($\pm$1$\sigma$ of the mean) in red. {\it
Bottom:} input spectrum, with the pixels selected for continuum
fitting indicated by red dots. The resulting continuum fit is
shown by the solid blue line.}\label{fitnoise}
\end{center}
\end{figure}

\begin{figure}
\begin{center}
\includegraphics[width=8cm]{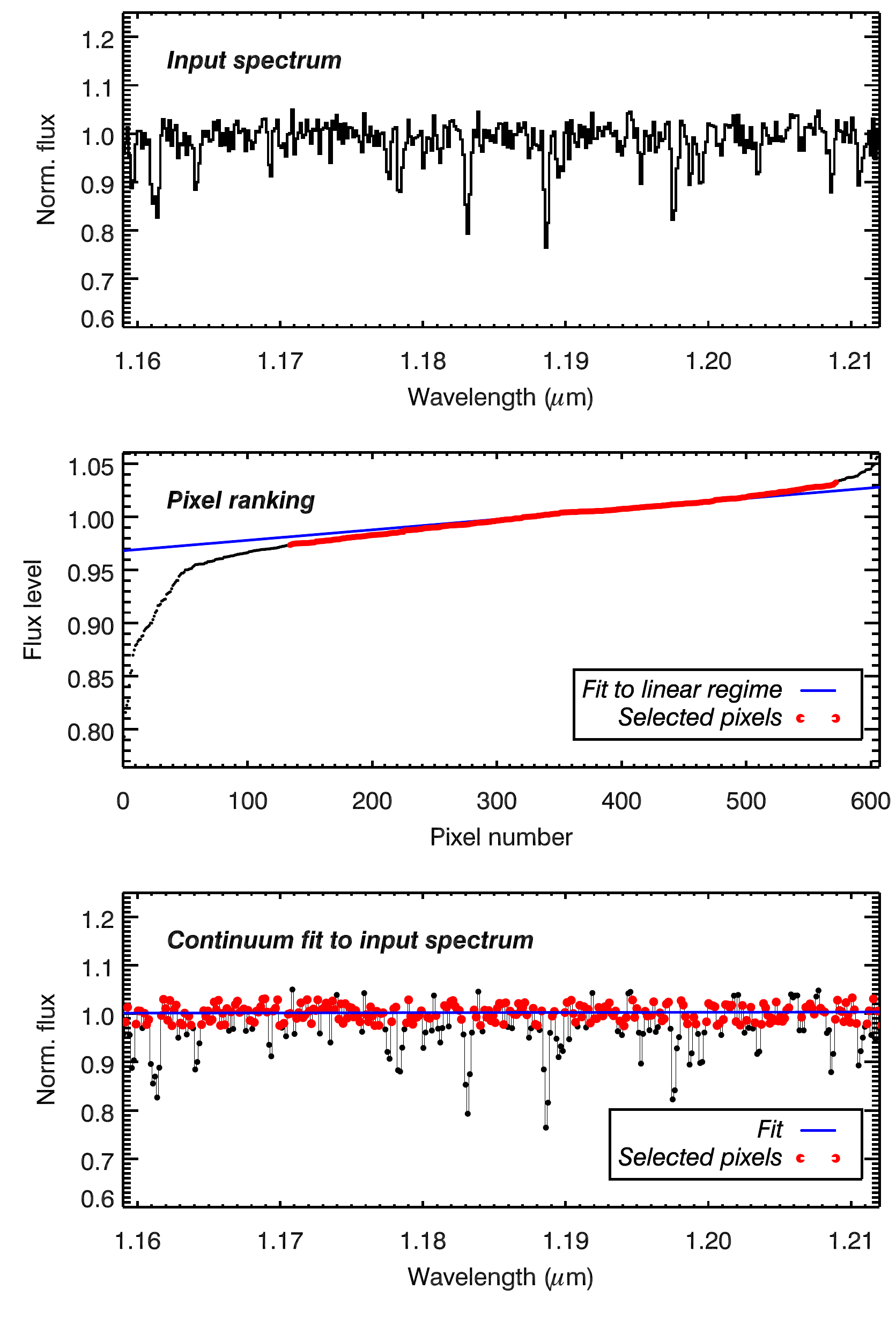}
\caption{As for Figure~\ref{fitnoise}, but now for the template
absorption-line spectrum, with Gaussian noised added
(such that S/N\,$=$\,50).}\label{fitspec}
\end{center}
\end{figure}

Here we adopt an improved method of continuum determination, assuming
that the noise is Gaussian (i.e. requiring good sky subtraction). A
spectrum of pure Gaussian noise, when ranked in order of increasing
pixel value, will produce a trend which is approximately linear in the
middle-ranking pixels (see Figure~\ref{fitnoise}). This linear regime
is selected to contain the values that are within $\sim$1$\sigma$ of
the mean.  An absorption-line spectrum (i.e. discrete features
combined with Gaussian noise) will also have this linear regime (see
Figure~\ref{fitspec}). By identifying and fitting these pixels, we can
find those within some tolerance level of the fit. In principle we can
then select those within $\sim$1$\sigma$ of the mean continuum level,
and use them to fit the continuum. Figures~\ref{fitnoise} and
\ref{fitspec} illustrate this procedure for a spectrum of pure
Gaussian noise and for a template absorption-line spectrum with noise
added (both with S/N\,$=$\,50).

\subsubsection{Refinement of the metallicity measurement}

The {\sc marcs} grid provides models at intervals of 0.25\,dex between
\logz\,=\,$-$1.5 and $+$1.0 (excepting $-$1.25). Determining a finer
estimate of the stellar metallicity requires that we interpolate
within the model grid.  DKF10 achieved this by taking a
\chisq-weighted mean of the best-fitting points in the grid. However,
if the \chisq\ trend with metallicity is asymmetric about the \chisq\
minimum (i.e.\ the best-fitting value), the `refined' metallicity will
always be systematically offset from the best-fitting value in the
direction of the shallowest \chisq\ gradient. For example, if the
\chisq\ minimum in the grid is at \logz\,=\,0.0, but the \chisq\
diverges more rapidly at negative \logz\ than at positive
values, the \chisq-weighted mean metallicity will always be above 0.0.
Indeed, in the course of the current study we found that determining
stellar metallicities in this way produced `refined' estimates that
were systematically larger than the input values (a consequence of the
morphology of the \chisq-minimum).

To eliminate these systematic errors we introduced a new method
to refine the metallicity determination beyond the resolution of the
model grid. After finding the location of the
\chisq\ minimum in the grid, we extract a two-dimensional plane in
\Teff-\logz\ space at equal \logg\ and $\xi$ values. We then resample
this grid onto a grid five times finer using bi-cubic spline
interpolation, and take the location of the \chisq\ minimum of the
resampled grid as the refined metallicity and stellar temperature. 

\subsection{Results}

Each simulated spectrum was run through the atmospheric analysis code
to determine \Teff, \logz, $\xi$ and \logg.  The mean metallicities
recovered (and their standard deviations) from the ten runs of each
simulation (for input metallicity, $R$, NGS configuration, seeing, and
effective magnitude) are shown in Figure~\ref{Zresults}.

\subsubsection{$R$\,$=$\,4000}
\begin{figure*}
\begin{center}
\includegraphics[scale=0.85]{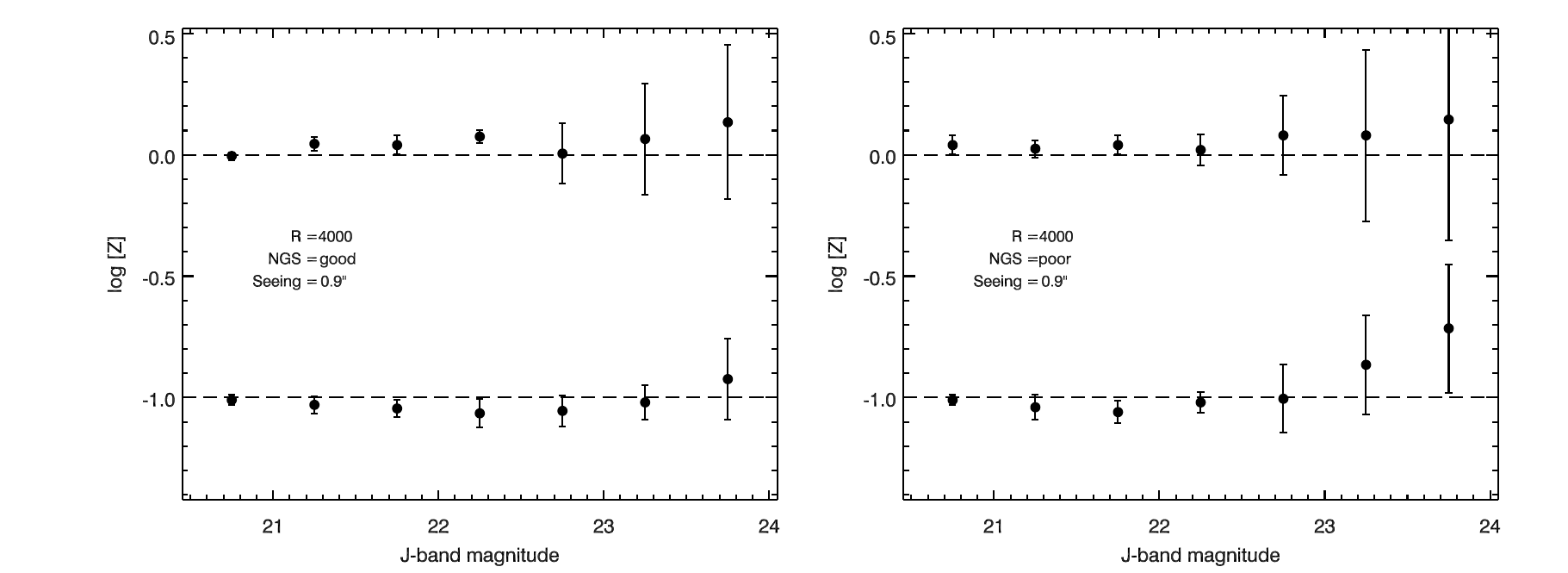}
\includegraphics[scale=0.85]{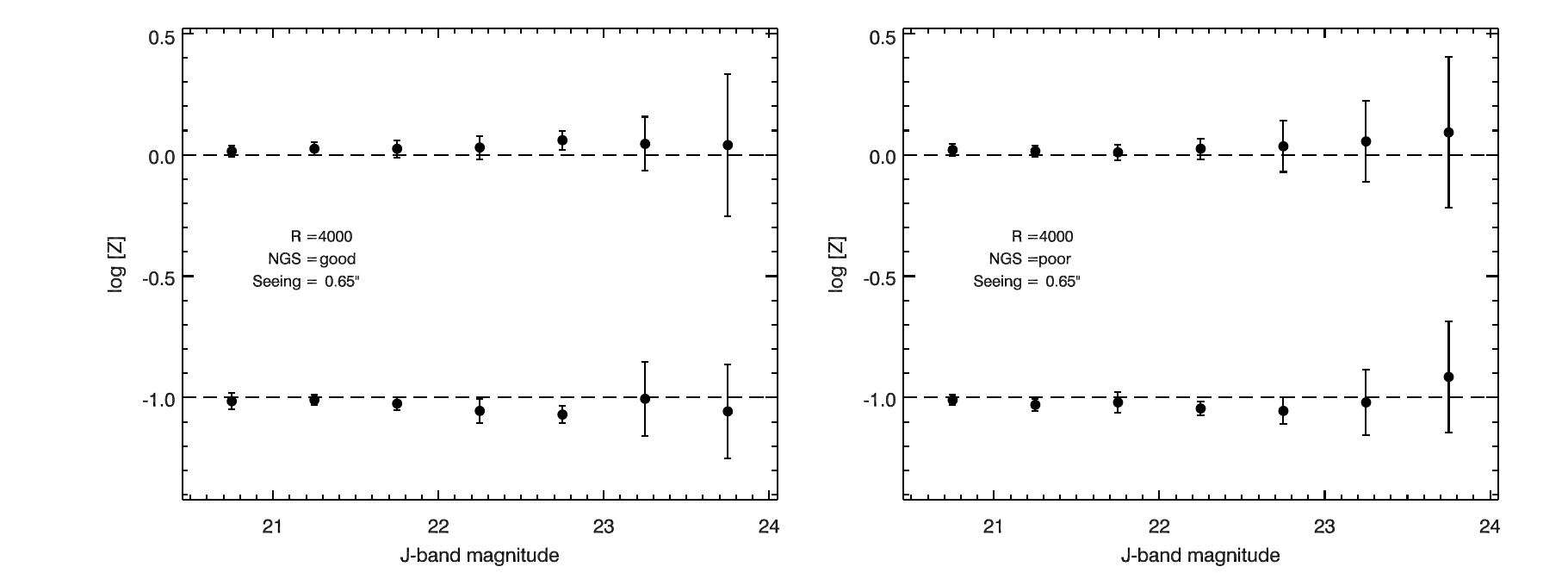}
\includegraphics[scale=0.85]{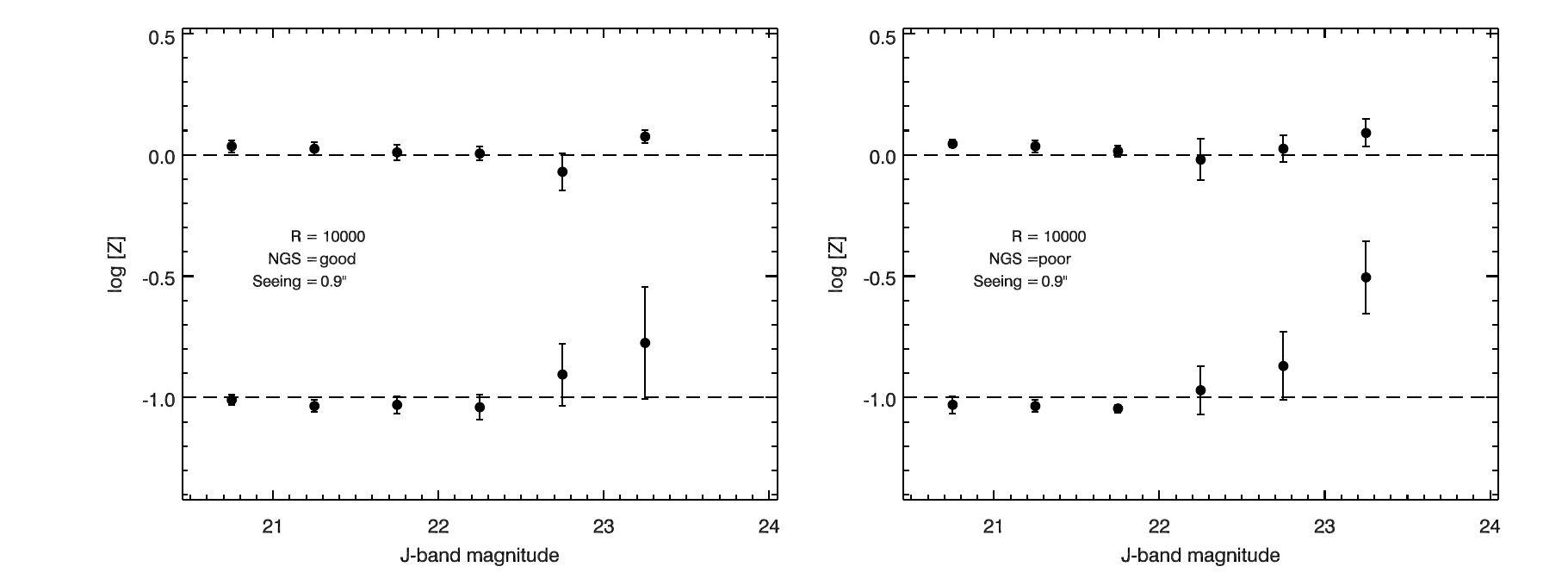}
\includegraphics[scale=0.85]{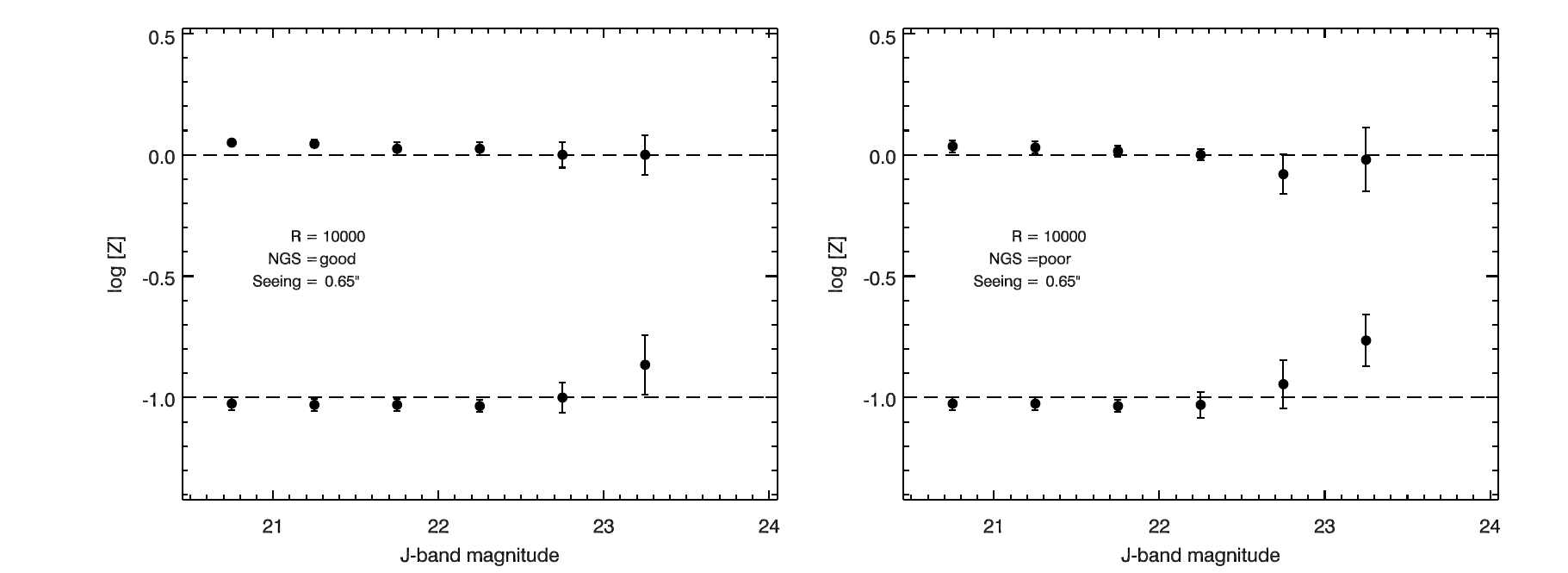}
\caption{Mean metallicities (and standard deviations) recovered from analysis 
of the simulated $J$-band spectra. Results are shown for the
\logz\,$=$\,0.0 and $-$1.0 input templates, at $R$\,$=$\,4000 and
10000, for both seeing values (0\farcs65 and 0\farcs9) and
configurations of natural guide stars (NGS; `good' and `poor').}\label{Zresults}
\end{center}
\end{figure*}

The mean metalliticites recovered from the $R$\,$=$\,4000 simulations
have a small scatter around the input value ($<$0.1\,dex) down to
$J$\,$=$\,22.75 (see Figure~\ref{Zresults}).  This is the point at
which the errors start to become larger, although the dispersion is
only $\pm$0.1-0.15\,dex for $J$\,$=$\,23.25 with 0\farcs65 seeing.  

These results provide a good test of the DKF10 methods for metal poor
stars, demonstrating reliable metallicity estimates at both
metallicities above a given S/N threshold.  Of course, S/N is not the
only effect which determines the accuracy of the metallicity estimate.
Other effects will likely play a role, such as the robustness of the
atomic data, any degeneracy between temperature, surface gravity, and
microturbulence, and which part of the curve-of-growth the selected
lines are on (i.e. the effects of saturation on the observed
equivalent width and its relation to abundance).  Nevertheless,
comparing these results with the S/N values in Table~\ref{Jsims}
suggests a minimum S/N of 50 to 60 is required to recover $Z$ to
within 0.1\,dex.

\subsubsection{$R$\,$=$\,10000}\label{results_10000}

Analysis of the $R$\,$=$\,10000 simulations also recovers good
metallicity determinations ($\Delta Z$\,$\pm$\,0.1\,dex) for both
templates. The minimum S/N threshold is marginally lower than for the
$R$\,$=$\,4000 spectra, with S/N\,$\gtrsim$\,45 required from comparison
with the results in Table~\ref{Jsims}.

Given the greater spectral dispersion, the faint magnitude limit is
(at least) half a magnitude brighter than for the $R$\,$=$\,4000
simulations.  Thus, use of the lower resolving power is (as to be 
expected) the most sensitive option if just a global estimate
of metallicity is required for a given target.  However, if precise
radial velocities are also required, to map the dynamics of a given
region of the host galaxy, then observations at $R$\,$=$\,10000
provide improved velocity resolution (30\,\kms\/, as compared to
75\,\kms\/ at $R$\,$=$\,4000).

To investigate the noise properties in the higher-resolution mode, the
intermediate files from the {\sc websim} were analysed.  The sky
background continuum counts ($\sim$60 photons/pixel for integrations
of 1800s) give a shot noise that is over three times greater than the
adopted detector read-noise.  The noise on the dark current is
actually the more critical factor in this regard, but note that our adopted
dark value is somewhat conservative \citep[e.g., a factor of two greater
than that measured for the arrays tested for {\em JWST}
by][]{figer04}. Shorter integrations of 900s remain background
limited, although minimising the detector read-noise in this instance
becomes an even more important requirement.

\subsection{$I$-band simulations}\label{results_I}

Understanding the relative performances for $I$- and $J$-band ELT
spectroscopy will be valuable when planning future observations.
Exploratory simulations of the CaT were given by \citeauthor{ao4elt}
(2010a) to investigate the potential of EAGLE spectroscopy in this
region.  To inform discussion of the new $J$-band results, we
revisited simulations of the CaT, now adopting the same {\sc marcs}
templates and seeing/NGS configurations as for the $J$-band
simulations.  The adopted
instrument throughput is slightly lower at 0.85\,$\mu$m (30\%)
compared to 1.175\,$\mu$m (35\%) in the final EAGLE design.  As
before, these values are not adopted to be definitive, but are
representative of a near-IR instrument solution in which the primary
performance requirements are at longer wavelengths.
Table~\ref{Isims} summarises the mean continuum S/N (and standard
deviations), per two-pixel resolution element, obtained from sets of
ten simulated spectra calculated at half-magnitude
intervals\footnote{The simulations were originally calculated adopting
the $I$-band zero-point from the {\em HST}-NICMOS conversion tools:
http://www.stsci.edu/hst/nicmos/tools/conversion$\_$help.html
(2250\,Jy at 0.9\,$\mu$m). Translated to the \citet{b79} system this
results in a shift of 0.1$^{\rm m}$ fainter, hence the sampling in
Table~\ref{Isims}.}.

As an external check on the simulated $I$-band performances we turn to
the ESO E-ELT spectroscopic exposure time calculator
(ETC)\footnote{http://www.eso.org/observing/etc/}.  Input parameters
were adopted to match those in the EAGLE simulations as closely as
possible: a {\sc marcs} model atmosphere with $T_{\rm
  eff}$\,$=$\,4000, point-source target, Paranal site, 42\,m primary,
seeing of 0\farcs8, zenith distance=30$^\circ$,
laser-tomography/multi-conjugate AO PSF, $R$\,$=$\,10000, a 20\,mas
radius of the S/N reference area (i.e. best match to our slice-width
of 37\,mas), and individual integration times of 1800\,s.  For
S/N\,$\ge$\,10 (per two-pixel resolution element), the ETC requires
10\,hrs of total integration time for $I$\,$=$\,24.85.  Note that the
simulated PSFs incorporated in the ESO ETC have some of the same
limitations to those generated for EAGLE, taking this into account
leads to an effective $I$-band magnitude of $\sim$23.5, i.e., of
comparable sensitivity to the results in Table~\ref{Isims}. Given the
differences in the AO architectures, the exact modelling tools (and
assumptions therein), and the methods used to calculate the S/N, the
ETC comparison provides an independent check that our sensitivity
limit (for a given S/N) is within the expected range.

\section{Discussion: RGB stars}\label{discussion} 

We now consider the simulation results in the context of the
(unreddened) distances to which we can recover accurate metallicities
from $J$- and $I$-band spectroscopy with the E-ELT.
For reference, absolute $I$-band magnitudes ($M_I$) for RGB stars span
$-$1\,$<$\,$M_I$\,$<$\,$-$3, with typical colours of ($I - J$)$_{\rm
  0}$\,$\sim$\,0.7-0.8 \citep[e.g.][]{g02,m08}.  At the tip of the
RGB absolute magnitudes extend up to $M_I$\,$=$\,$-$4.

\subsection{Potential for E-ELT \& EAGLE}

We require S/N\,$\gtrsim$\,50 for accurate metallicities from the
$J$-band.  At $R$\,$=$\,4000, the results in Table~\ref{Jsims} provide
sufficient S/N down to $J$\,$=$\,22.75 to 23.25, depending on the NGS
configuration and seeing.  Taking the bright end of the RGB (adopting
$M_J$\,$=$\,$-$3.75 from the colours above), this corresponds to an
unreddened distance modulus of 27, or a distance of $\sim$2.5\,Mpc.

Thus, the DKF10 methods could provide a direct metallicity indicator
in RGB stars out to a volume which encompasses a diverse range of
galaxies.  For instance, as we move beyond the Local Group, there are
32 known galaxies with distances in the range 1.0\,$<$\,d\,$\le$\,2.65\,Mpc
\citep{k04}.  Most notably these include NGC\,55 and NGC\,300, the two
spirals in the closer part of the Sculptor `Group' at 1.9\,Mpc
\citep{gieren05,gieren08,p06}.  NGC\,55 is seen to have
`LMC-like' abundances from spectroscopy of luminous supergiants
\citep[][\& in prep]{castro08} and NGC\,300 has a radial oxygen
gradient that ranges from 0.1\,dex below Solar to near LMC levels at a
radius of $\sim$5\,kpc \citep[][and references therein]{b09}.  This
range of distances also includes lower-mass galaxies such as
Sextans~A, NGC\,3109, and GR\,8, which are all depleted in metals (as
traced by oxygen) by approximately 1\,dex
\citep[][respectively]{kaufer04,e07,vsh06}.  Of course, some galaxies
within 2.5\,Mpc, in addition to the faint dwarfs in the Local Group,
will have even lower metallicities, providing motivation to further
explore these diagnostics in the very metal-poor regime.

Scaling these results for TRGB stars ($M_J$\,$\sim$\,$-$5), good
metallicities should be recoverable out to distances of 4 to 5\,Mpc,
depending on the exact AO correction.  This opens-up an even wider
range of targets, encompassing over 200 galaxies \citep[][]{k04},
including those in the Centaurus~A Group \citep[e.g.][]{k02}.  Direct
metallicities for TRGB stars will play an important role in calibration of the
distance scale in the local universe, reducing the uncertainties
associated with photometric distance determinations from the TRGB method
\citep[see discussion by][]{kud10}, which can lead to 
systematic uncertatinies of $\sim$5\% \citep[e.g.][]{mmf08}.

\subsection{Comparison with $I$-band performances}

To compare the $J$-band results with the commonly used CaT diagnostic,
we now consider the $I$-band performances given in Table~\ref{Isims}.
To determine the limiting magnitude for metallicities from our
$I$-band simulations, we adopt a requirement of S/N\,$\ge$\,20 (per
two-pixel resolution element).  \citet{bit08} consider S/N\,$\ge$\,20
(per \AA) as the threshold to obtain metallicities to within 0.1\,dex
from CaT spectroscopy at $R$\,$=$\,6500 (from VLT-FLAMES).  At the
centre of our $I$-band template 1\,\AA\/ is equivalent to 2.33 pixels,
so the S/N results in Table~\ref{Isims} are slightly conservative in
this regard, but they are also at greater spectral resolving power.
However, the dominant factor here is the lower AO performance in the
$I$-band than at longer wavelengths.  The EE in the simulated PSFs is
already considerably lower than in the $J$-band (Table~\ref{psfs}),
but the additional error terms are also larger (i.e. a factor of four
compared to a factor of two in the $J$-band).

Depending on the seeing conditions and NGS configuration, the $I$-band
simulations yield S/N\,$\ge$\,20 for 22.1\,$<$\,$I$\,$<$\,23.1,
comparable to the $J$-band limit discussed in the previous section
(S/N\,$\gtrsim$\,50 at $R$\,$=$\,4000).  Given that RGB/TRGB stars are
moderately red ($I - J$\,$=$\,0.6-1.0), the DKF10 technique is probing
deeper in terms of absolute luminosity of a given star.

The $J$-band technique exploits the improved AO correction at longer
wavelengths. It requires greater S/N to recover a good estimate of
stellar metallicity but, given the final EAGLE instrument throughputs
and current expectations of AO performance shortwards of 1\,$\mu$m,
outperforms the CaT observations for a fixed exposure time of a given
RGB star.  In addition to a {\em direct} metallicity measurement
(rather than the CaT calibration, albeit thought to be well
understood), $J$-band observations will have the advantage of reduced
extinction compared to observations at shorter wavelengths.  Although
we have limited ourselves to consideration of extra-galactic sources,
working at longer wavelengths would be attractive for observations in,
for example, obscured Galactic clusters.

\subsection{Potential for TMT-IRMS}
Our simulations can also be considered in the context of the Infra-Red
Multi-Slit Spectrometer (IRMS) being developed for the TMT
\citep{s10}, which is a clone of the MOSFIRE instrument for the
Keck\,I telescope.  IRMS will be located behind the TMT
multi-conjugate AO system (NFIRAOS), taking advantage of excellent
image quality across a 2$'$ field.  Its planned wavelength coverage is
0.9-2.5\,$\mu$m and, depending on the slit-widths, the delivered
resolving powers will be $R$\,$=$\,3000 to 5000.  

The minimum (Nyquist sampled) IRMS slit-width will be 160\,mas, i.e.
coarser spatial sampling and with a greater sky contribution for a
point-source target than from EAGLE. The expected $J$-band {\em
encircled} energies for a slit-width of 160\,mas range from
approximately 30 to 45\,\%, depending on the position within the 2$'$
field (Dr.~Brent Ellerbroek, private communication).  These
simulations include some of the additional error terms factored into
the discussion of our MOAO PSFs, i.e., when the larger IRMS
slit-widths are taken into account, they provide a comparable range of
image quality.

Assuming that the sensitivity varies as a factor of primary aperture
(combined with a greater background flux from a larger effective
slit), this corresponds to observations of RGB stars out to distances
of $\sim$1.5\,Mpc and stars at the TRGB out to $\sim$3\,Mpc (depending
on limitations of the NFIRAOS performance estimates cf. those
discussed in Section~\ref{moaopsfs}). Tailored performance simulations
are warranted for stellar spectroscopy specific to the TMT design,
instrumentation and site (selected to be Mauna Kea in Hawaii).
Nevertheless, we highlight the potential of the DKF10 methods for TMT
spectroscopy of stellar populations beyond the Local Group.  For
example, in NGC\,1569, a northern starburst galaxy at 1.95\,Mpc
\citep{k04} with a metal abundance that appears to be intermediate to
those of the LMC and SMC \citep[\citeauthor{drd97} 1997,
\citeauthor{ks97} 1997, cf., for example,][]{t07}.  The TMT will also
be well placed to explore the wealth of faint targets in northern
hemisphere galaxies that are closer to home, e.g., M31 and M33.

\section{Discussion: Red supergiants}\label{rsgs}

The work by DKF10 was originally motivated by spectroscopy of RSGs.
With the benefit of AO correction and their intrinsically red colours,
ELT observations of RSGs will provide our most distant quantitative
tracers of individual `normal' stars. Extending the earlier discussion, and with
absolute magnitudes of $-$7\,$<$\,$M_J$\,$<$\,$-$11, $J$\,$=$\,23.25
(at $R$\,$=$\,4000) corresponds to an impressive 70\,Mpc for the most
luminous RSGs. The potential of such observations requires
investigation in greater detail. In particular, the impact of crowding
(and the required AO correction) will need to be assessed, for
instance, using the {\em HST} observations of galaxies in the Virgo
and Fornax Clusters \citep{cote04,j07}.

\subsection{Potential for VLT-KMOS}
More iminently, these results are relevant for KMOS (under
construction for the VLT), which will have 24 IFUs, each
2\farcs8\,$\times$\,2\farcs8 on the sky (with 0\farcs2/slice),
providing $R$\,$=$\,3400 in the {\em YJ} bands \citep{kmos10}.  The
ETC for the natural-seeing mode of SINFONI provides the most
appropriate comparison available at present -- the limiting magnitude
for S/N\,$\sim$\,50, in a total exposure time of 10\,hrs, is
$J$\,$=$\,18.75.  As noted by DKF10, this will enable us to obtain
surveys of RSGs in galaxies well beyond the Local Group, and for the
brightest targets at distances of up to $\sim$10\,Mpc. This will
provide the first opportunity to obtain large samples of RSGs to
investigate large-scale metallicity gradients and structure in
external galaxies.

\subsection{Potential for {\em JWST}-NIRSpec}

These methods will also have applications with the {\em JWST} Near-IR
Spectrometer (NIRSpec), which is a multi-object spectrograph capable
of observing over 100 sources over a field of approximately
3.5\,$\times$\,3.5 arcmin.  NIRSpec will have three selectable
spectral resolving powers ($R$\,$=$\,100, 1000 \& 2700), over a range
of 1-5\,$\mu$m.

The expected point-source NIRSpec
sensitivity\footnote{http://www.stsci.edu/jwst/instruments/nirspec/}
for S/N\,$=$\,10 (two-pixel resolution element) with the
$R$\,$=$\,2700 grating, at 1.15\,$\mu$m, is 950.7\,nJy (equating to
$J_{\rm VEGA}$\,$=$\,23.1) in a total integration of 100\,ks
(27.8\,hr).  This suggests a limiting magnitude of $J$\,$=$\,19.6
(for S/N\,$\sim$\,50) which, combined with the large multiplex of the
mult-slit array, could obtain large samples of RSGs in galaxies out
to the Virgo Cluster.

\section{Summary}

We have simulated $J$-band EAGLE/E-ELT observations of an M0 giant
(and, by extrapolation of the model atmospheres, an M0 supergiant).
We extended the work of DKF10 for the 1.15-1.22\,$\mu$m diagnostic
region, finding:
\begin{itemize}
\item{Robust stellar metallicities can be obtained from analysis of $J$-band spectra of 
metal-poor stars (\logz\,$=$\,$-$1.0).}
\item{Accurate metallicities ($\pm$0.1\,dex, at \logz\,$=$\,$-$1.0 and 0.0) require:
S/N\,$\ge$\,55 at $R$\,$=$\,4000, and S/N\,$\ge$\,45 at $R$\,$=$\,10000.}
\item{The Phase~A design of EAGLE on a 42\,m E-ELT has the potential
    of {\em direct} metallicity estimates for RGB stars out to
    distances of $\sim$2.5\,Mpc, and for RSGs well beyond the Virgo
    and Fornax Clusters (subject to crowding).}
\item{For our instrument assumptions and simulations of relative AO
    performance, the $J$-band diagnostics are slightly more sensitive
    than the CaT ($\Delta$ mag $\sim$0.5-1.0) for recovering
    metallicities of a given M0-type star.  The $J$-band method also
    has the advantage of reduced extinction compared to the CaT
    region.}
\end{itemize}

The first two of these points are not specific to EAGLE (nor dependent
on the simulated PSFs used here), serving as a proof-of-concept for
broader application of the methods advanced by DKF10. This part of the
$J$-band is relatively free of OH emission lines compared to other
parts of the near-IR, enhancing its potential for quantitative
spectroscopy of extragalactic stellar populations with ELTs.  This
diagnostic range also has great potential for other upcoming
facilities such as VLT-KMOS, {\em JWST}-NIRSpec, and TMT-IRMS.

One of the prime drivers for inclusion of the $I$-band in the
design of near-IR spectrographs for ELTs is to study extragalactic
stellar abundances via the CaT. If comparable or better results can be
obtained from the $J$-band, inclusion of the $I$-band would become
less critical, leading to simpler and potentially cheaper instrument
designs (e.g. fewer gratings), and avoiding compromises which may
detract from the potential performance at longer wavelengths.

As noted by DKF10, an important next step is quantitative analysis of
RSGs in the Magellanic Clouds to refine these techniques over a range
of metallicities, and to also extend this work to RGB stars in the
Clouds.  These methods should also be explored at yet lower
metallicities (relevant to the study of metal-poor galaxy halos), both
via analysis of metal-poor observations, and by use of a larger grid
of theoretical calculations.

Diagnostics in the $H$- and $K$-bands (in which the AO correction will
be even better) also warrant consideration in this context.  For
instance, use of $K$-band equivalent-width indices to estimate stellar
metallicities \citep{f01}, and detailed abundance analysis of red
giants \citep{ors02,or04}, and supergiants \citep{d09a,d09b} at
greater resolving powers in the $H$-band.

\vspace{0.2in}
\noindent{\sc acknowledgments:} We thank Fran\c{c}ois Ass\'{e}mat
for his work on the EAGLE AO study, Brent Ellerbroek for updated information
on the TMT simulations, and the referee for their
constructive comments on the manuscript.  BD is funded by a fellowship
from the Royal Astronomical Society. RPK acknowledges support from the
Alexander-von-Humboldt Foundation.  MP, JGC and GR acknowledge support from the
Agence Nationale de la Recherche under contract ANR-06-BLAN-0191.  DF
is supported by NASA under award NNG 05-GC37G, through the Long Term
Space Astrophysics program, and by a NYSTAR Faculty Development
Program grant.

\bibliography{Jspec}
\bibliographystyle{aa} 

\end{document}